\documentclass{article} 

\usepackage{graphicx}
\usepackage{amsmath,amssymb}

\begin{document}

\title{Incorporating Cellular Stochasticity in Solid--Fluid Mixture Biofilm Models}

\author{Ana Carpio, Universidad Complutense de Madrid, Spain \\
            Elena Cebrian, Universidad de Burgos, Spain}
            
\date{Jan 31, 2020}            

\maketitle

{\bf Abstract.} The dynamics of cellular aggregates is driven by the interplay of mechanochemical processes and cellular activity. Although deterministic models may capture mechanical features,  local chemical fluctuations trigger random cell responses, which determine the overall evolution. Incorporating stochastic cellular behavior in macroscopic models of biological media is a challenging task. Herein, we propose hybrid models for bacterial biofilm growth, which couple a two phase solid/fluid mixture description of mechanical and chemical fields with a dynamic energy budget-based cellular automata treatment of bacterial activity. Thin film and plate approximations for the relevant interfaces allow us to obtain numerical solutions exhibiting behaviors observed in experiments, such as accelerated spread due to water intake from the environment, wrinkle formation, undulated contour development, and the appearance of  inhomogeneous distributions of differentiated bacteria performing varied tasks.
A single paragraph of about 200 words maximum. For research articles, abstracts should give a pertinent overview of the work. We strongly encourage authors to use the following style of structured abstracts, but without headings: (1) Background: Place the question addressed in a broad context and highlight the purpose of the study; (2) Methods: Describe briefly the main methods or treatments applied; (3) Results: Summarize the article's main findings; and (4) Conclusion: Indicate the main conclusions or interpretations. The abstract should be an objective representation of the article, it must not contain results which are not presented and substantiated in the main text and should not exaggerate the main conclusions. 


\section{Introduction}

Bacterial biofilms provide basic model environments for analyzing the 
interaction between mechanical and cellular aspects of three-dimensional 
self-organization during development. Biofilms are formed when bacteria encase 
themselves in a hydrated layer of self-produced extracellular matrix (ECM) made 
of exopolymeric substances (EPS) \cite{flemming}. This habitat confers them 
enhanced resistance to disinfectants, antibiotics, flows, and other mechanical or 
chemical agents \cite{hoiby}.

Research on modeling biofilms has increased steadily during the past few 
decades resulting in the understanding of a number of features. Continuous
models for uniform cell distributions are useful in basic culture systems
\cite{stewart}. Individual based models \cite{ib1,ib2} and cellular automata 
\cite{ca} may capture variable thickness, density, and structure. However, 
current models focus more on deterministic mass transfer and extracellular 
structure, than in random cell processes. Interest on fluctuations in 
intracellular concentrations, for instance, has arisen due to their significance
in phenotypic variability as well as in gene regulation and stochasticity of
gene expression \cite{sto1,sto2}, with consequences for development
and drug resistance \cite{birnir}.

\begin{figure}[!hbt]
\centering
\includegraphics[width=8cm,angle=0]{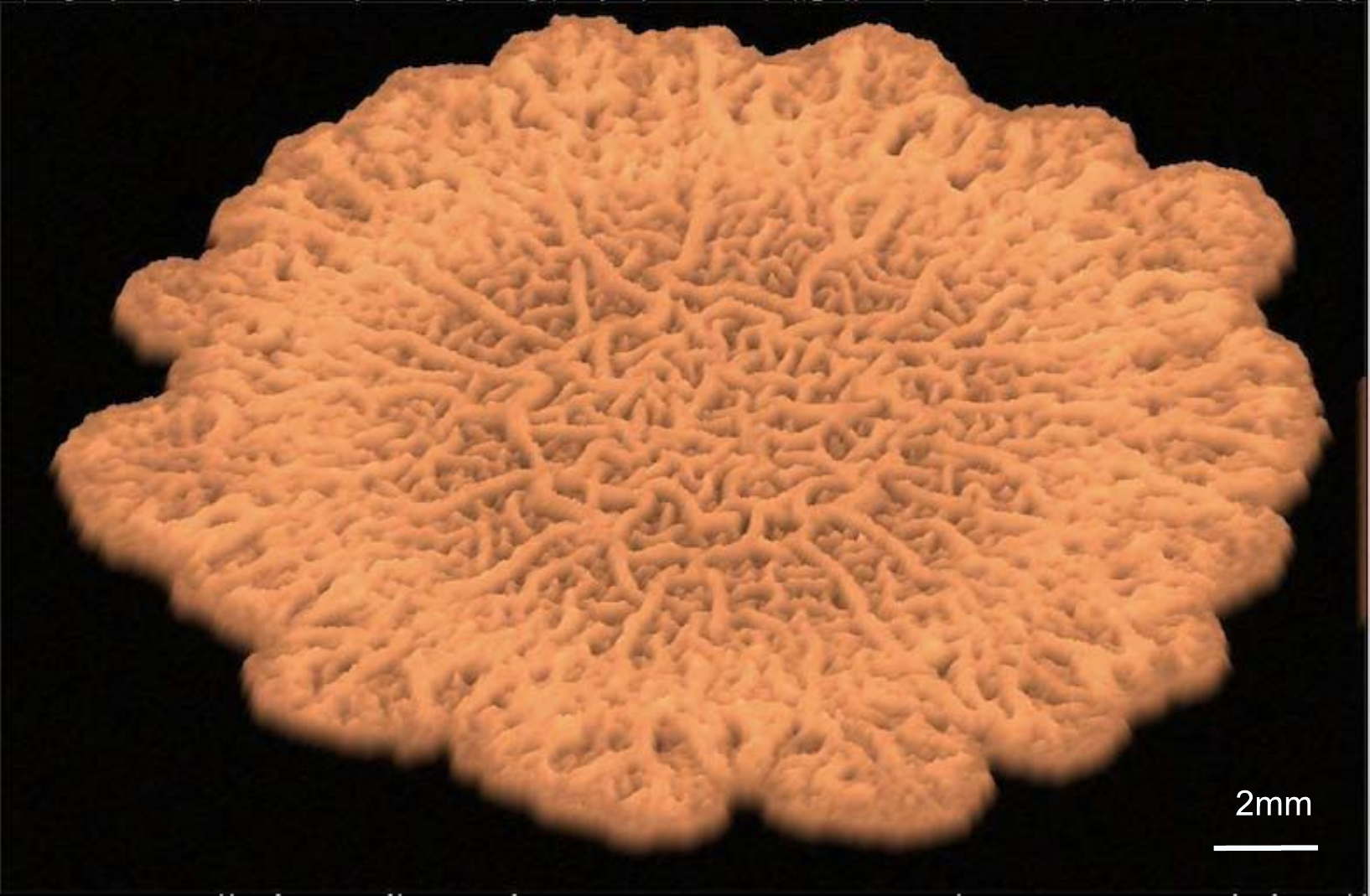}  
\caption{Virtual visualization of a biofilm spreading on agar.}
\label{fig1}
\end{figure}

{Recent experiments with \textit{Bacillus subtilus} biofilms on agar provide 
a case study in which we can test models incorporating new aspects.
Once bacteria adhere to a surface, they differentiate in response to 
local fluctuations created by growth, death, and division processes, to 
variations in the concentrations of nutrients, waste, and autoinducers,
to cell--cell communication \cite{hera}.
Some of them become producers of exopolymeric substances (EPS) and
form the extracellular matrix (ECM). EPS production increases the osmotic
pressure in the biofilm, driving water from the agar substrate and 
accelerating   spread~\cite{seminara}. In addition, the matrix confers
the biofilm elastic properties. Wrinkles develop as the result of
localized death in regions of high cell density
and compression caused by division and growth \cite{asally}. As~the biofilm 
expands, complex wrinkled patterns develop, see Figure~\ref{fig1}. 
This~phenomenon is linked to gradients created by heterogeneous cellular 
activity and water migration~\cite{espeso}. Eventually, the wrinkles form a 
network of channels transporting water, nutrients, and waste to sustain it 
\cite{wilking,wingreen}.
Biofilm spread due to osmosis can be accounted for by two-phase flow 
models and thin film approximations~\cite{seminara}. Instead,
wrinkle formation has been reproduced by means of  Von K\'arm\'an-type theories { \cite{espeso, wrinkles}}. 
Delamination and folding processes are further analyzed  in \cite{benamar1} 
by  means  of neo-Hookean models. In~\cite{poroelastico},
a~poroelastic approach provides a unified description of liquid transport
and elastic deformations in the biofilm. To incorporate fluctuations in a more 
natural way, here we propose a mixture model allowing to distinguish the 
different phenotypes forming the film.}

{Biofilm structure is greatly influenced by environmental
conditions. When they grow in flows, we find bacteria  immersed in large 
lumps of polymer, typically forming fingers and streamers \cite{stone}
in the surrounding current. In contrast, biofilms spreading on air--agar interfaces 
contain small volume fractions of extracellular matrix  \cite{seminara}, 
producing wrinkled shapes with internal water flow. This motivates 
different treatments of the extracellular matrix, see \cite{ib0, ib3} for biofilms 
in flows and \cite{ib1,ib2,seminara,basic1} for biofilms on interfaces with
air or tissues, for instance. In the latter case, when internal fluid flow is taken
into account, the small fraction of matrix is usually merged in one biomass 
phase with the cells \cite{seminara,basic1}. Some experimental studies 
suggest a viscoelastic rheology for biofilms \cite{viscoelastic1,viscoelastic2}.
The analysis of the mixture and poroelastic models we consider  
shows that, depending on the volume fractions of solid biomass and fluid, the 
viscosity of the fluid, the Lam\'e constants of the solid, the densities, and
hydraulic permeability of the fluid/solid system, the characteristic
time for variations in the displacement of the solid, and the characteristic
length of the network in the macroscopic scale, the resulting 
mixture can be considered as monophasic elastic, monophasic 
viscoelastic, or truly biphasic mixture/poroelastic \cite{keller,kapellos}.}

The  paper is organized as follows. Section \ref{sec:mixture}
introduces the solid--fluid mixture model. Section~\ref{sec:stochastic}
discusses ways to incorporate details of cell behavior. We present
a cellular automata approach based on dynamic energy budget
descriptions of bacterial metabolism.
With the aid of asymptotic analysis \cite{seminara, witelski}, 
we construct numerical solutions  displaying behaviors consistent with 
experimental observations.
Finally, Section \ref{sec:discussion} discusses our results, 
{the advantages, and limitations of our approach, as well 
as future perspectives and possible improvements.}

\section{Solid-Fluid Mixture Model of a Biofilm Spreading on an Agar/Air  Interface}
\label{sec:mixture}

\begin{figure}[!hbt]
\centering
\hskip -4cm  (\textbf{a}) \hskip 5cm (\textbf{b}) \\
\includegraphics[width=10cm,angle=0]{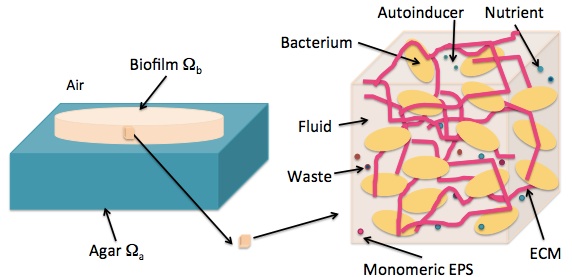}  
\caption{Schematic structure of a biofilm: 
(\textbf{a}) View of the macroscopic configuration: a biofilm on an agar--air interface.
(\textbf{b}) Microstructure formed by biomass (polymeric mesh and cells) and 
fluid containing dissolved substances (nutrients, waste, and autoinducers).}
\label{fig2}
\end{figure}

{In this section,} we adapt  bicomponent mixture models of 
swelling tissues \cite{basic1,basic2} to describe the spread of biofilms
of air--agar interfaces, including biomass variations.
We consider the biofilm as a bicomponent mixture of incompressible 
solid matrix (bacterial cells and polymers) and interstitial fluid carrying 
nutrients, waste, and autoinducers, see Figure \ref{fig2}.
The biofilm occupies a region $\Omega_b(t),$ placed over an agar 
substratum $\Omega_a(t)$ and in contact with air. 
Large immobilized solutes are considered part of the extracellular
matrix (ECM). Small molecules diffusing rapidly are considered part of 
the fluid. 

\subsection{Mass Balance}
\label{sec:lcmass}

Under the equipresence hypothesis of mixtures, each point $\mathbf x$ in 
the biofilm can be occupied simultaneously by both phases. In addition, 
we assume that no air bubbles or voids form inside the biofilm. If 
$\phi_s(\mathbf x,t) $ denotes the volume fraction of solid and 
$\phi_f(\mathbf x,t) $ the volume fraction of fluid, then $\phi_s + \phi_f = 1.$
The standard densities of biological tissues and agar are similar to the density of 
water $\rho_f = 10^3 \, {\rm kg}/{\rm m}^3$ (typical relative differences of order 
$10^{-2}$). Thus, we will take the densities of all constituents and the mixture to 
be constant and equal to that of water \cite{seminara}: $\rho_f= \rho_s= \rho= \rho_w.$
Then, the balance laws for the fractions of solid biomass $\phi_s$ and fluid 
$\phi_f$ are  \cite{basic1, seminara}
\begin{eqnarray} 
{\partial \phi_s \over \partial t} + {\rm div}  (\phi_s {\mathbf v}_s) = 
 r_s(\phi_s,c_n),  \quad  
{\partial \phi_f \over \partial t} + {\rm div} (\phi_f {\mathbf v}_f) 
= - r_s(\phi_s,c_n), 
\label{lcsolid}  
\end{eqnarray}
where  $r_s(\phi_s,c_n)$ represents  biomass production due to
nutrient consumption, whereas $\mathbf v_s$ and $\mathbf v_f$ stand 
for the velocities of the solid and the fluid components, respectively.
Biomass production can be accounted for through a Monod law 
$
r_s(\phi_s,c_n) =  k_s {c_n \over c_n + K_n} \phi_s = 
{1+\alpha_m \over \tau} {c_n \over c_n + K_n} \phi_s =g(c_n) \phi_s,
$ 
where $c_n$ is the nutrient concentration, $K_n$ is a constant
that marks the onset of starvation, and $k_s$ the uptake rate, 
which can be approximated by ${1+\alpha_m \over \tau}$,
$\tau$ being the doubling time for the specific bacteria and 
$\alpha_m$ a factor representing polymeric matrix production
\cite{seminara}. 

Adding up Equation (\ref{lcsolid}), we find a conservation law for the 
growing mixture:
\begin{eqnarray}
0= {\rm div}  (\phi_s {\mathbf v}_s+ \phi_f {\mathbf v}_f) = 
{\rm div}  (\mathbf v) = {\rm div} (\mathbf v_s + \mathbf q),
\label{balancemixture}
\end{eqnarray}
where $\mathbf v= \phi_s {\mathbf v}_s+ \phi_f {\mathbf v}_f$ is the
composite velocity of the mixture and
\begin{eqnarray}
\mathbf q = \phi_f (\mathbf v_f - \mathbf v_s) = \phi_f \mathbf w
\label{flux}
\end{eqnarray}
is the filtration flux, $\mathbf w$ being the relative velocity.

\subsection{Driving Forces}
\label{sec:forces}

Forces inducing motion are of a different nature: inner stresses, inertial forces, 
interactive forces, and external body forces. We discuss here the constitutive 
relations and  fluxes for incompressible solid--fluid mixtures in the case of 
infinitesimal deformations and under isothermal conditions \cite{basic1}.

\subsubsection{Stresses in the Solid and the Fluid}
\label{sec:stresses}

When a large number of small pores are present in the biofilm, the 
stresses in the fluid are  
\begin{eqnarray}
\boldsymbol \sigma_f =  - \phi_f \, p \mathbf I,
\label{stressfluid}
\end{eqnarray}
where $p$ is the pore hydrostatic pressure. In the presence of large regions 
filled with fluid, the overall fluid shear stresses should be considered too.
The stresses in the solid arise from the strain within the solid and from the 
interaction with the fluid. Under small deformations, and for an isotropic solid, 
we~have
\begin{eqnarray}
\boldsymbol \sigma_s =  \hat{\boldsymbol \sigma}_s  - \phi_s \, p \mathbf I, 
\quad \hat{\boldsymbol \sigma}_s = \lambda_s {\rm Tr} 
(\boldsymbol  \varepsilon({\mathbf u}_s))\, \mathbf I +  
2 \mu_s \, \boldsymbol  \varepsilon({\mathbf u}_s), \label{stresssolid} \\[1ex]
\varepsilon_{ij}({\mathbf u})= {1\over 2} \Big(  {\partial u_i \over \partial x_j } 
+ {\partial u_j \over \partial x_i}  \Big), \nonumber
\end{eqnarray}
where ${\mathbf u}_s$  is the displacement vector of the solid biomass;
$ \boldsymbol  \varepsilon({\mathbf u})$ the deformation tensor;
and $\lambda_s, \mu_s,$ the Lam\'e constants, related to the
Young $E$ and Poisson $\nu$ moduli by
$\lambda = {E \nu \over (1+\nu) (1-2\nu)}, $ $
\mu =  {E \over 2 (1+\nu)}. $

\subsubsection{Interaction and Inertial Forces}
\label{sec:interaction}

In most biological samples, the velocities $\mathbf v_f$ and $\mathbf v_s$ 
are small enough for inertial forces to be negligible: 
$\rho_s \mathbf a_s \approx \rho_f \mathbf a_f \approx \rho \mathbf a \approx 0$, 
where $\mathbf a_s, \mathbf a_f, \mathbf a$ represent the solid, fluid, and composite accelerations. Thus, we will work in a quasi-static deformation regime.

The interaction forces act on the two components. They are opposite in sign 
and equal in magnitude, as a result, their combined effect vanishes on the tissue. 
We consider two kinds: filtration resistance and concentration
gradients in chemical potentials.

The filtration resistance arises from the interaction between fluid and solid particles. 
Per unit volume, these forces are  $\phi_s \mathbf f_s$, 
$\phi_f \mathbf f_f$ and satisfy $\phi_s \mathbf f_s+\phi_f \mathbf f_f =0.$
In the absence of inertial effects and concentration--viscous couplings 
$\mathbf f_f= -\boldsymbol \alpha \mathbf q$, where $\boldsymbol \alpha(\phi_f)$
is the resistivity matrix and $\mathbf q$ the filtration flux. For isotropic elastic 
solids with isotropic permeability
\begin{eqnarray}
\mathbf f_f = - {1\over k_h} \mathbf q,
\label{filtrationforce}
\end{eqnarray}
where $k_h={k \over \mu_f}$ is the hydraulic permeability, $k$ being the 
permeability of the solid, and $\mu_f$ the fluid viscosity.  Typically,
$k_h(\phi_f)={\phi_f^2 \over \zeta}$, where $\zeta$ is a friction
parameter often taken to be $\zeta= {\mu_f \over \xi(\phi_s)^2} >0$
and $\xi$ represents the ``mesh size'' of the matrix network.

The concentration forces in the fluid $\nabla \pi_f$ are
$\nabla \pi_f = - {1\over \hat V_f} \nabla \mu^{f,c}, $ 
where $\hat V_f$ is the molar volume of the fluid and 
$\mu^{f,c}$ is the concentration contribution to the chemical  
potential of the fluid $\mu^{f}$. Under isothermal conditions
\begin{eqnarray}
\nabla \mu^f = \hat V_f \nabla p - \nabla \mu^{f,c}
= \hat V_f \nabla (p  - \pi_f).
\label{chemical}
\end{eqnarray}
Similar relations hold for concentration forces $\nabla \pi_s$ in 
the solid, which satisfy $\phi_s \nabla \pi_s  + \phi_f \nabla \pi_f = 0.$

\subsection{Equations of Motion}
\label{sec:motion}

The theory of mixtures hypothesizes that the motion of each constituent
is governed by the usual balance equations, as if it was isolated 
from the other one. Neglecting inertial terms, and in the absence of
external body forces, the momentum balance for the solid and the
fluid reads
\begin{eqnarray*}
{\rm div} \boldsymbol \sigma_s + \phi_s (\mathbf  f_s + \nabla \pi_s) = 0,\quad 
{\rm div} \boldsymbol \sigma_f  + \phi_f  (\mathbf  f_f + \nabla \pi_f) = 0.
\end{eqnarray*}

Using the expressions for the stress tensors (\ref{stressfluid}) and (\ref{stresssolid}), 
these equations become
\begin{eqnarray}
{\rm div} \, \hat{\boldsymbol \sigma}_s  
+ \phi_s (-\nabla p + \nabla \pi_s ) + \phi_s \mathbf  f_s = 0,
\quad
\phi_f  (-\nabla p+\nabla \pi_f) + \phi_f  \mathbf  f_f = 0.
\label{motionsolidfluid}
\end{eqnarray}

Combining (\ref{motionsolidfluid}), (\ref{filtrationforce}), and (\ref{flux})
we obtain the Darcy law in the presence of concentration 
gradients
\begin{eqnarray}
\mathbf q = - k_h \nabla (p - \pi_f) = \phi_f (\mathbf v_f -\mathbf v_s).
\label{darcy}
\end{eqnarray}

Adding up equations (\ref{motionsolidfluid}), we find an equation relating solid 
displacements and pressure
\begin{eqnarray}
{\rm div} \, \hat{\boldsymbol \sigma}_s(\mathbf u_s)
- \nabla p = 0.
\label{motionsolid2}
\end{eqnarray}

The solid velocity is then $\mathbf v_s ={\partial \mathbf u_s
\over \partial t}.$ These equations are complemented by the conservation
of mass (\ref{lcsolid}) and (\ref{balancemixture}), which now reads as
\begin{eqnarray}
{\rm div}(\mathbf v_s) = - {\rm div}(\mathbf q) =
{\rm div}(k_h \nabla (p - \pi_f)).
\label{balancemixture2}
\end{eqnarray}

The flux (\ref{darcy}) can be rewritten as 
$\mathbf q = k_h {\rm div}(- \hat{\boldsymbol \sigma}_s + \pi_f \mathbf I),$
where $- \hat{\boldsymbol \sigma}_s + \pi_f \mathbf I$ is the swelling
stress.
In biphasic swelling theory \cite{basic2}, it is customary to work with
$p- \pi_f = p_f$, where $p_f$ is associated to the fluid chemical potential
by (\ref{chemical}) and $\pi_f$ is identified with the osmotic pressure
created by the concentration of a specific chemical 
\cite{basic2,ghassemi}.  
The osmotic pressure in the biofilm is caused by
the concentration of EPS produced by the cells and can be taken
to be proportional to the volume fraction of solid biomass $\pi_f =
\Pi \phi_s,$ $\Pi$ being the osmotic compressibility \cite{seminara}.
Equation (\ref{motionsolid2}) then motivates the introduction of
effective constitutive laws for the whole mixture of the form \cite{basic2}
$\boldsymbol \sigma (\mathbf u)=
\hat{\boldsymbol \sigma}_s(\mathbf u) - p \mathbf I
= \hat{\boldsymbol \sigma}_s(\mathbf u) - (p_f+\pi_f) \mathbf I,$
as usual in poroelastic theory.

\subsection{Final  Equations}
\label{sec:final}

Summarizing the main governing equations, we get
\begin{eqnarray}
{\partial \phi_s \over \partial t} + {\rm div}  (\phi_s {\mathbf v}_s) 
= g(c_n) \phi_s,  & \label{finalphis} \\ [1ex]  
{\rm div} (\mathbf v_s ) = {\rm div}(k_h(\phi_f)
\nabla p_f), & 
\label{finaldarcy} \\ [1.5ex]
\mu_s \Delta \mathbf u_s + (\mu_s + \lambda_s) 
\nabla ({\rm div}(\mathbf u_s)) =  \nabla (p_f + \pi_f), &
\label{finalus} \\
{\mathbf v_s} ={\partial \mathbf u_s \over \partial t}, \quad
\pi_f=\pi(\phi_s) \label{finalvs},
\end{eqnarray}
in the region occupied by the biofilm $\Omega_b(t)$, which
varies with time.
In equilibrium, $\mathbf q = \mathbf v_s = \mathbf f_f=0.$
At the biofilm boundary, the jumps in the total
stress vector  and  the chemical potential vanish:
\begin{eqnarray*}
(\hat{\boldsymbol \sigma}_s  - p \mathbf I) \mathbf n =
(\hat{\boldsymbol \sigma}_s  - (p_f+\pi_f) \mathbf I) \mathbf n 
= \mathbf t_{ext},  \quad
p_f =  p - \pi_f = p_{ext} - \pi_{f,ext}, 
\end{eqnarray*}
when applicable.
In this quasi-static framework, the displacements $\mathbf u_s$ 
depend on time through the motion of the biofilm boundary. 

If we need to track the variations of the nutrient concentration, we
may use effective continuity equations for chemical concentration in 
tissues  \cite{chen,porotissue}. For the limiting concentration
$c_n$:
\begin{eqnarray}
{\partial  c_n \over \partial t} + {\rm div}  (\mathbf v_f c_n)  
- {\rm div} (d_n \nabla c_n)   =  -r_n(\phi_s,c_n), \quad
r_n(\phi_s, c_n) \!=\! \phi_s k_n { c_n \over c_n + K_n},
\label{lcnutrient} 
\end{eqnarray}
where $d_n$ is an effective diffusivity \cite{effectiveconcentration}.
Setting $d_{n,s}$ and $d_{n,f}$, the diffusivities in the biomass
and liquid  $d_n = d_{n,f} {3 d - 2 \phi_f (d-1) \over
3 + \phi_f (d-1)}$, $d=K_{eq} {d_{n,s} \over d_{n,f}}.$
The  source $r_n(\phi_s,c_n)$ represents consumption by the biofilm,
$k_n$ being the uptake rate and $K_n$ the half-saturation constant.
Zero-flux boundary conditions are imposed at the air--biofilm interface.
Instead, at the agar--biofilm interface, we may impose a constant 
concentration through a Dirichlet boundary condition. Being more 
realistic, we couple this diffusion equation to another one defined in 
the agar substratum $\Omega_a(t)$ with zero source and transmission 
conditions at the interface \cite{espeso}.

\begin{figure}[!hbt]
\centering
 \hskip -4cm  (\textbf{a}) \hskip 5cm (\textbf{b}) \\
 \includegraphics[width=10cm,angle=0]{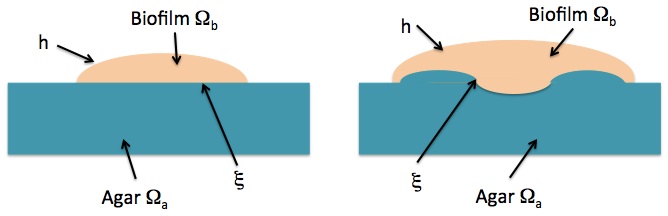}  
\caption{Schematic representation of a biofilm slice, with moving
air--biofilm interface $h$ and agar--biofilm interface $\xi$.
(\textbf{a})  Initial stages: $\xi$ is flat.
(\textbf{b})  Later evolution: $\xi$ deviates out of a plane. }
\label{fig3}
\end{figure}

Solving Equations (\ref{finalphis})--(\ref{finalvs}), studying the evolution of a 
biofilm also requires tracking the dynamics of the biofilm boundary, see Figure \ref{fig3}.  
In principle, there are two boundaries of a different nature: the air--biofilm interface 
and the agar--biofilm interface.  

\subsection{Motion of the Air--Biofilm Interface}
\label{sec:airbiofilm}

During the first stages of biofilm spread,  the agar--biofilm interface remains 
flat, whereas the biofilm reaches a height $x_3=h(x_1,x_2,t)$, see Figure 
\ref{fig3}a. Integrating~(\ref{balancemixture}) in the $x_3$ direction we obtain
$
\begin{array}{l}
\int_{0}^h {\partial (\mathbf v \cdot \hat{\mathbf x}_1) \over \partial x_1} \, dx_3 + 
\int_{0}^h {\partial (\mathbf v \cdot \hat{\mathbf x}_2) \over \partial x_2} \, dx_3 + 
\int_{0}^h {\partial (\mathbf v \cdot \hat{\mathbf x}_3) \over \partial x_3} \, dx_3 
=0, \end{array}
$
$\hat{\mathbf x}_1$, $\hat{\mathbf x}_2$ and $\hat{\mathbf x}_3$ being the 
unit vectors in the Cartesian coordinate directions. By Leibniz's rule 
$
\begin{array}{l}
\int_{0}^h {\partial (\mathbf v \cdot \hat{\mathbf x}_i) \over \partial x_i} \, dx_3
= {\partial \over \partial x_i} \left[
\int_{0}^h (\mathbf v \cdot \hat{\mathbf x}_i) \, dx_3 \right]
- \mathbf v \cdot \hat{\mathbf x}_i\big|_h {\partial h \over \partial x_i}, 
\quad i=1,2.
\end{array}
$
Therefore,
\begin{eqnarray}
\begin{array}{l}
{\partial \over \partial x_1} \left[
\int_{0}^h (\mathbf v \cdot \hat{\mathbf x}_1) \, dx_3 \right] + 
{\partial \over \partial x_2} \left[
\int_{0}^h (\mathbf v \cdot \hat{\mathbf x}_2) \, dx_3 \right]
- \mathbf v \cdot \hat{\mathbf x}_1\big|_h {\partial h \over \partial x_1} 
\\[2ex]
- \mathbf v \cdot \hat{\mathbf x}_2\big|_h {\partial h \over \partial x_2}
+ \mathbf v \cdot \hat{\mathbf x}_3\big|_h  =
\mathbf v \cdot \hat{\mathbf x}_3\big|_0.
\end{array} 
\label{height1}
\end{eqnarray}

Differentiating $x_3(t)=h(x_1(t),x_2(t),t)$ with respect to time and
using $\mathbf v \cdot \hat{\mathbf x}_i = {dx_i \over dt}$, $i=1,2,3,$ we find
$
\mathbf v \cdot x_3 \big|_h = {d x_3 \over dt} = {d \over dt} h(x_1(t),x_2(t),t) 
= {\partial h \over \partial t} + {\partial h \over \partial x_1} {d x_1 \over dt}
+ {\partial h \over \partial x_2} {d x_2 \over dt}   
= {\partial h \over \partial t} + 
\mathbf v \cdot x_1 \big|_h  {\partial h \over \partial x_1} 
+ \mathbf v \cdot x_2 \big|_h {\partial h \over \partial x_2} .
$
Inserting this identity in (\ref{height1}), we find the equation
\begin{eqnarray}
\begin{array}{l}
{\partial h \over \partial t } +
{\partial \over \partial x_1} \left[
\int_{0}^h (\mathbf v \cdot \hat{\mathbf x}_1) \, dx_3 \right]
+ {\partial \over \partial x_2} \left[
\int_{0}^h (\mathbf v \cdot \hat{\mathbf x}_2) \, dx_3 \right]  =
\mathbf v \cdot \hat{\mathbf x}_3\big|_0,
\end{array} 
\label{height2}
\end{eqnarray}
where $\mathbf v \cdot \hat{\mathbf x}_i =
{d u_{s,i} \over dt} - k_h(\phi_f)  {\partial p_f \over \partial x_i}, i=1,2,3.$
To obtain a closed equation for the height $h$ we need to calculate the 
velocity of the solid $\mathbf v_s={d \mathbf u_{s} \over dt}$, the modified 
pressure $p_f$ and the volume fraction of fluid from  
(\ref{finalphis}) and (\ref{finaldarcy}). 
This equation is able to describe accelerated spread due to osmosis, 
at least in simplified  geometries, as we illustrate next.

From Equation (\ref{height2}), we derive an approximated equation for the 
early evolution of the height of a circular biofilm, see Appendix \ref{sec:apA} 
for details and assumptions
\begin{eqnarray} \begin{array}{l} \displaystyle
h_t \!-\! {K R\over R_0} {e^{3t} \over r} (r h_{rt} h^3 )_r 
\!-\! {3K R\over 2 R_0} {e^{3t}  \over r} (r h_{r} h^2 h_t )_r  
\!-\!  {K R\over R_0}  {e^{3t}  \over r} (r h_{r} h^3 )_r  \\[2ex]
\displaystyle
\!-\! {3K R\over 2R_0}  {e^{3t}  \over r} (r h_{r} h^3 )_r
\!-\! {K R_t\over R_0} {e^{3t } \over r} (r h_{r} h^3 )_r  \!=\! 0.
\end{array}
\label{echeight} 
\end{eqnarray}

A simplified version 
\begin{eqnarray}
h_t  - K(1+{3\over 2}) R  e^{3t }{1\over r} (r h_{r} h^3 )_r = 0, \quad
K={ {g \mu_f}  \over 3 \xi_\infty^2 \mu_s (1-\phi_\infty)^2 R_0}  {h_0^3},
\label{echeightseminara}
\end{eqnarray}
has self-similar solutions.  Restoring dimensions, they take the
form
\begin{eqnarray} \begin{array}{l}
\displaystyle
h= h_0 e^{gt} (R/R_0)^{-2} f({r \over R}) = e^{gt} (R/R_0)^{-2} 
 (1- {3\over 2} \, {r^2 \over R^{2}})^{1\over 3}, \\
 \displaystyle
R=  R_0 \left({7 \over 3} K(1+{3\over 2}) (e^{3gt }-1) +1
\right)^{1 \over 7}.
\end{array} \label{selfsimilarhR}
\end{eqnarray}

Replacing $(1+3/2)$ by $1$ in (\ref{selfsimilarhR}), we recover
the self-similar solution found in \cite{seminara}, with $g \mu_f$
instead of $\mu_f$ ($\mu_f$ being the fluid viscosity) and
the Lam\'e coefficient of the solid biomass $\mu_s$ instead of  
the  viscosity of the fluid biomass $\mu_s$.

\begin{figure}[!hbt]
\centering
\includegraphics[width=8cm,angle=0]{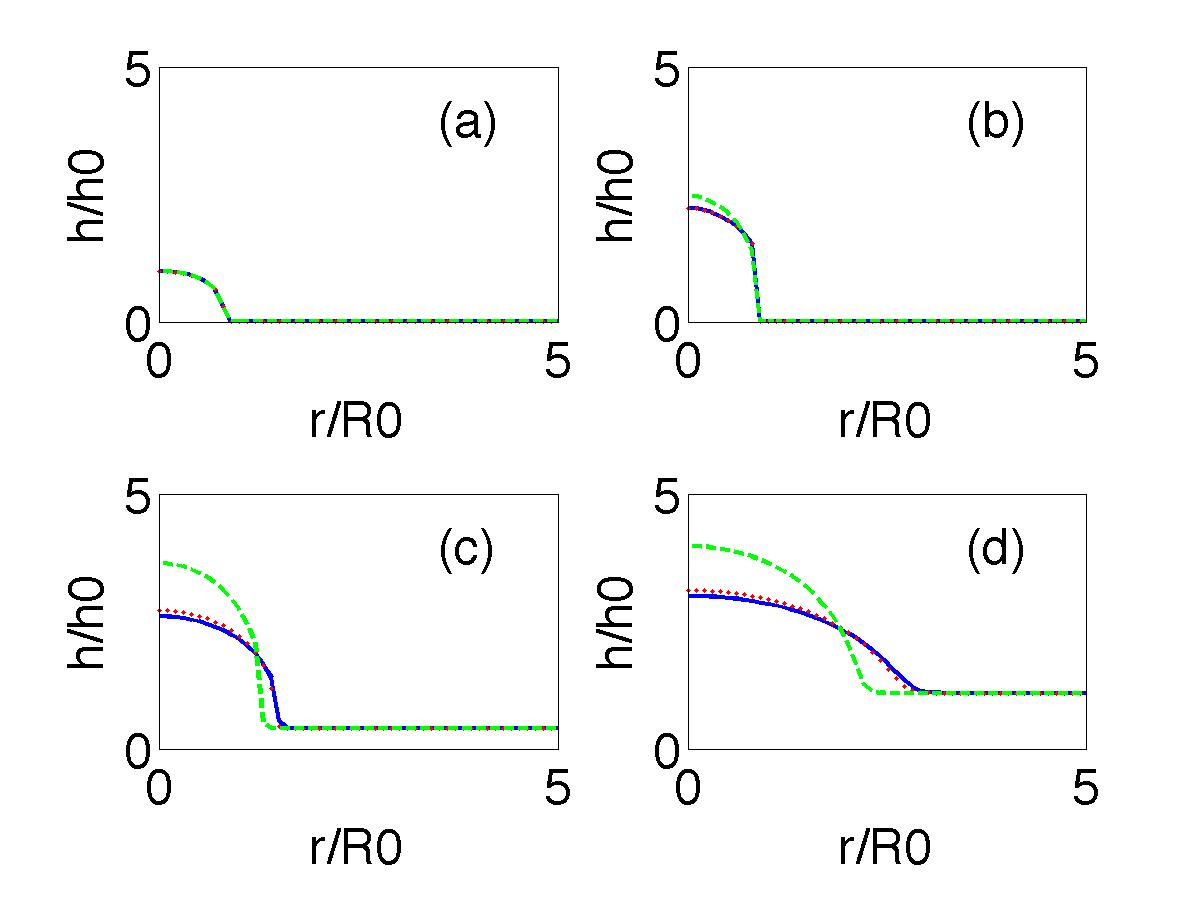}  
\caption{Biofilm height at dimensionless times $0$ (\textbf{a}), $1$ (\textbf{b}), $6$ 
(\textbf{c}), and $7$ (\textbf{d}) for $K=10^{-5}$ and $h_{\inf}=10^{-3}.$ 
The dotted red line and the solid blue line depict the numerical 
solutions of  (\ref{echeightseminara}) and (\ref{echeight}),
respectively, with $R$ given by (\ref{selfsimilarhR})  and keeping 
the same data. We can observe the transition from an initial
stage in which increase in biofilm height dominates to a 
stage with faster horizontal spread. The green line is a reference
self-similar approximation.}
\label{fig9}
\end{figure}

Figure \ref{fig9} compares the time evolution of the 
biofilm height profiles starting from a smoothed version of
(\ref{selfsimilarhR}). Notice that (\ref{selfsimilarhR}) only makes
sense when $R^2> 3/2 \, r^2$, and that the slope diverges 
at $r = \sqrt{2/3} R$. Experiments show that a thin biofilm
layer precedes the advance of the biofilm bulk \cite{seminara}.
We set $h=h_{\infty}>0$ beyond that point. The dashed
green line in Figure \ref{fig9} represents the numerical
solution of (\ref{echeightseminara}), with $R$ given by
(\ref{selfsimilarhR}) for $K=10^{-5}$, and $h_{\inf}=10^{-3},$
replacing $(1+3/2)$ by $1$ as in \cite{seminara}.
The dotted red line and the solid blue line depict the numerical 
solution of  (\ref{echeightseminara}) and (\ref{echeight}),
respectively, with $R$ given by (\ref{selfsimilarhR})  and keeping 
the same data. They all show the transition from vertical growth 
to horizontal spreading as time goes on. The effect of the 
additional time derivatives in (\ref{echeight}) is to flatten the
profiles.

When  the interface biofilm/agar is not flat, but admits a parametrization of 
the form $x_3=  \xi(x_1,x_2,t)$,  as in Figure \ref{fig3}b,  
\begin{eqnarray}\begin{array}{l}
\int_{\xi}^h {\partial (\mathbf v \cdot \hat{\mathbf x}_i) \over \partial x_i} \, 
dx_3 = {\partial \over \partial x_i} \left[
\int_{\xi}^h (\mathbf v \cdot \hat{\mathbf x}_i) \, dx_3 \right]
- \mathbf v \cdot \hat{\mathbf x}_i\big|_h {\partial h \over \partial x_i}
+  \mathbf v \cdot \hat{\mathbf x}_i\big|_{\xi} {\partial \xi \over \partial x_i}, 
\quad i=1,2.
\end{array} \nonumber \end{eqnarray}
Repeating the previous computations in the interval $[\xi,h]$, the equation 
for the biofilm height becomes
\begin{eqnarray}
\begin{array}{l} \displaystyle
{\partial h \over \partial t } +
{\partial \over \partial x_1} \left[
\int_{\xi}^h (\mathbf v \cdot \hat{\mathbf x}_1) \, dx_3 \right]
+ {\partial \over \partial x_2} \left[
\int_{\xi}^h (\mathbf v \cdot \hat{\mathbf x}_2) \, dx_3 \right]   
=  \\ [1ex]
\displaystyle
\mathbf v \cdot \hat{\mathbf x}_3\big|_\xi
- \mathbf v \cdot \hat{\mathbf x}_1\big|_{\xi} {\partial \xi \over \partial x_1}
- \mathbf v \cdot \hat{\mathbf x}_2\big|_{\xi} {\partial \xi \over \partial x_2}.
\end{array} 
\label{height3}
\end{eqnarray}

Knowing $\xi$, this equation can be solved numerically coupled to
(\ref{finalphis}) and (\ref{finaldarcy}).

\subsection{Motion of the Agar/Biofilm Interface}
\label{sec:agarbiofilm}

Equations for the dynamics of the agar--biofilm interface follow using a 
Von Karman-type approximation, as the thickness of the biofilms
is small compared to its radius. Although  initially flat,
the displacements in the direction orthogonal to the interface may 
become large. Thus, the linear definition of the strain and stress tensors 
in  (\ref{stresssolid}) is replaced by \cite{landau}
\begin{eqnarray}
\varepsilon_{i,j}&=&{1\over 2} \Big( {\partial u_{i} \over \partial x_{j}} 
+ {\partial u_{j} \over \partial x_{i}}  
+ {\partial \xi \over \partial x_{i}} {\partial \xi \over \partial x_{j}}
\Big) + \varepsilon_{i,j}^0, \quad i=1,2,
\label{strain}
\end{eqnarray}
which includes nonlinear terms, as well as residual strains $\varepsilon_{i,j}^0$.  
We denote the  in-plane displacements by ${\bf u}=(u_1(x_1,x_2,t),u_2(x_1,x_2,t))$ 
and the out-of-plane displacements of the interface by $\xi(x_1,x_2,t)$. 
The coordinates $(x_1,x_2)$ vary along the 2D 
projection of the  3D biofilm structure on the biofilm/agar interface.
Equation (\ref{finalus}) becomes ${\rm div} \hat {\boldsymbol \sigma}
= \nabla (p_f + \pi_f)$, with $\hat {\boldsymbol \sigma}$ given by
(\ref{stresssolid}) and (\ref{strain}). 
{Formally, this allows us to identify
the biofilm with an elastic film growing on a viscoelastic agar substratum.
The~pressure terms become residual stresses.}
Then, the interface motion is governed by the equations~\cite{huang,espeso}:
\begin{eqnarray}
{\partial \xi \over \partial t} = {1 - 2 \nu_a \over 2 (1-\nu_a)} {h_a \over \eta_a}
\Big[ D (- \Delta^2 \xi + \Delta C_M)  
+ h {\partial\over \partial x_{j}} \left( \sigma_{i,j}({\bf u}) 
{\partial \xi \over \partial x_{i}}\right) \Big]
-{\mu_v \over \eta_a} \xi, \label{plategrowth1bis} \\
{\partial {\bf u} \over \partial t} = {h_a h \over \eta_a} 
{\rm div}({\boldsymbol \sigma({\bf u})}) - {\mu_v \over \eta_a} {\bf u}, 
\label{plategrowth2bis}  
\end{eqnarray}
where $h_a$ is the thickness of the viscoelastic agar substratum and 
$\mu_v$, $\nu_a$, and $\eta_a$  its rubbery modulus, Poisson ratio, and  
viscosity, respectively. The tensor $\boldsymbol \sigma$ is given by
\begin{eqnarray}
\begin{array}{l}
\sigma_{11}= {E \over 1-\nu^2} (\varepsilon_{11}+ \nu \varepsilon_{22})
+ \sigma_{11}^0,  \quad
\sigma_{12}= {E \over 1+\nu} \varepsilon_{12} + \sigma_{12}^0, \\[1ex]
\sigma_{22}= {E \over 1-\nu^2} (\varepsilon_{22}+ \nu \varepsilon_{11})
+ \sigma_{22}^0, 
\end{array} \label{stress} 
\end{eqnarray}
with $\boldsymbol \varepsilon$ defined in (\ref{strain}); $\nu$
and $E$ being the Poisson and Young moduli of the biofilm
(\ref{stresssolid}), respectively; {and $\boldsymbol \sigma^0$
represents the residual stresses.}
The bending stiffness is $D={E h^3\over 12(1-\nu^2)}$,  
$h$ being the initial biofilm thickness.
Here, the first Equation (\ref{plategrowth1bis}) describes out-of-plane bending  
$\xi$, and the second one (\ref{plategrowth2bis}) governs in-plane stretching 
for the displacements ${\bf u}=(u_1,u_2)$. Modified equations taking into 
account possible spatial variations in the elastic moduli are given in \cite{sergei}.

To identify the relevant scales governing the evolution of the agar--biofilm
interface we nondimensionalize (\ref{plategrowth1bis}) and (\ref{plategrowth2bis}). 
Making the change of variables
$\hat {\bf x}= {{\bf x} \over R},$ $ \hat {\bf u}= {{\bf u} \over R}, $ $
\hat \xi= {\xi \over h}, $ $ \hat \sigma={\sigma \over E}, $ $ \hat t 
= {t \over T},$
where $R$ is the approximate biofilm radius, and setting $R=\gamma h$, 
the dimensionless equations become
\begin{eqnarray}
{\partial \hat \xi \over \partial \hat t} =
\Bigg[  {12(1-\nu^2) \gamma^2}  {\partial\over \partial \hat x_{j}}
 \left( \hat \sigma_{i,j}(\hat{\bf u}) {\partial \hat \xi \over \partial \hat x_i}\right) 
 + (- \Delta_{\hat{\bf x}}^2 \hat \xi + \Delta_{\hat{\bf x}} \hat C_M) \Bigg]  
- T {\mu_a \over \eta_a} \hat \xi, \label{plategrowth1adim} \\
{\partial \hat{\bf u} \over \partial \hat t} = \tau \;
{\rm div}_{\hat{\bf x}}{\hat {\boldsymbol \sigma}(\hat{\bf u}) } 
- T {\mu_a \over \eta_a} \hat {\bf u}, \label{plategrowth2adim} 
\end{eqnarray}
where $ T =  {2 (1-\nu_v) \over 1-2 \nu_v}{\eta_v h \over h_v} {12(1-\nu^2) 
\gamma^4\over E } = \tau {\eta_v h \over h_v E} \gamma^2, $ $
\tau = 24 {(1-\nu_v)\over (1-2\nu_v) } (1-\nu^2)\gamma^2. $
Wrinkled structures develop when the nonlinear terms are large
enough, therefore $\gamma={R\over h}$ must be large enough.

\begin{figure}[!hbt]
\centering
 (\textbf{a}) \hskip 5.5cm (\textbf{b})   \\ 
\includegraphics[height=4.5cm]{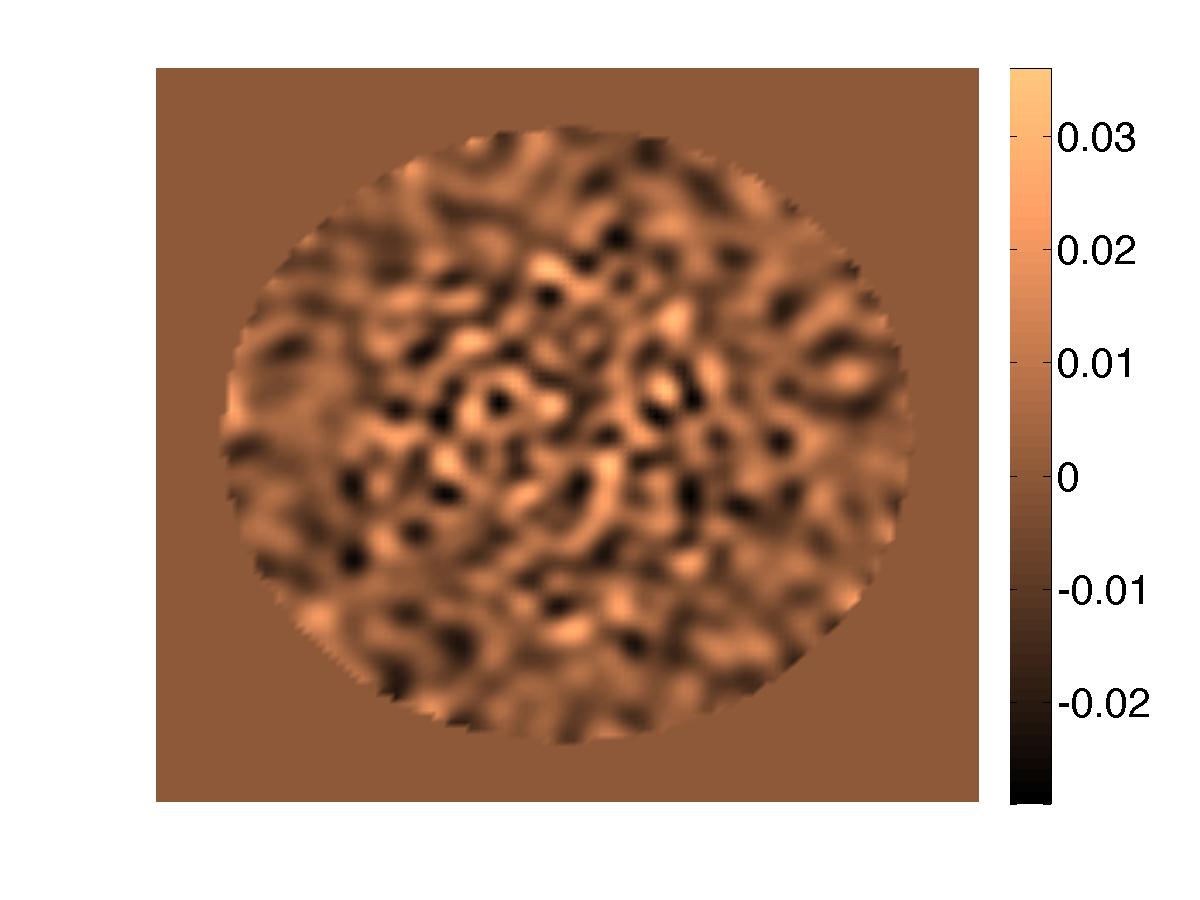}   
\includegraphics[height=4.5cm]{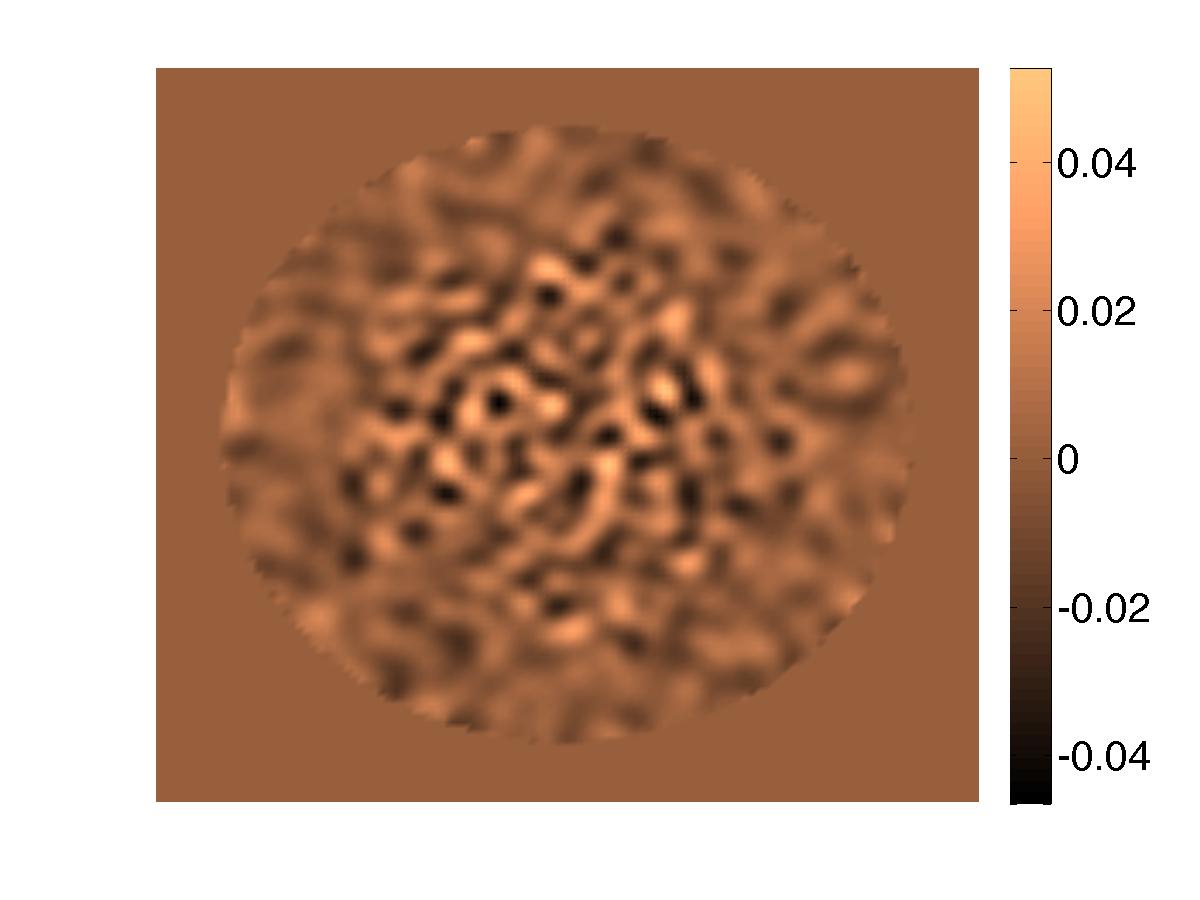}   \\
(\textbf{c}) \hskip 5.5cm (\textbf{d}) \\
\includegraphics[height=4.5cm]{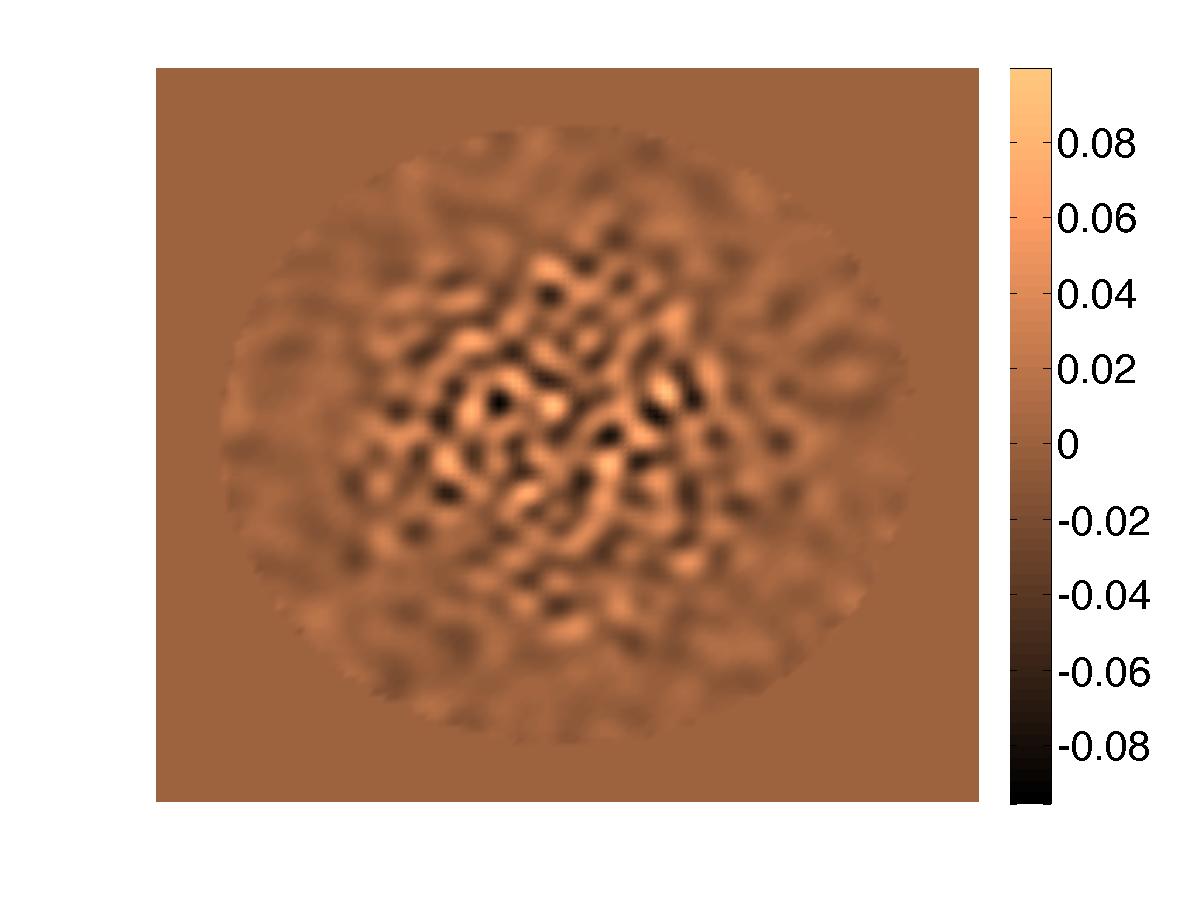} 
\includegraphics[height=4.5cm]{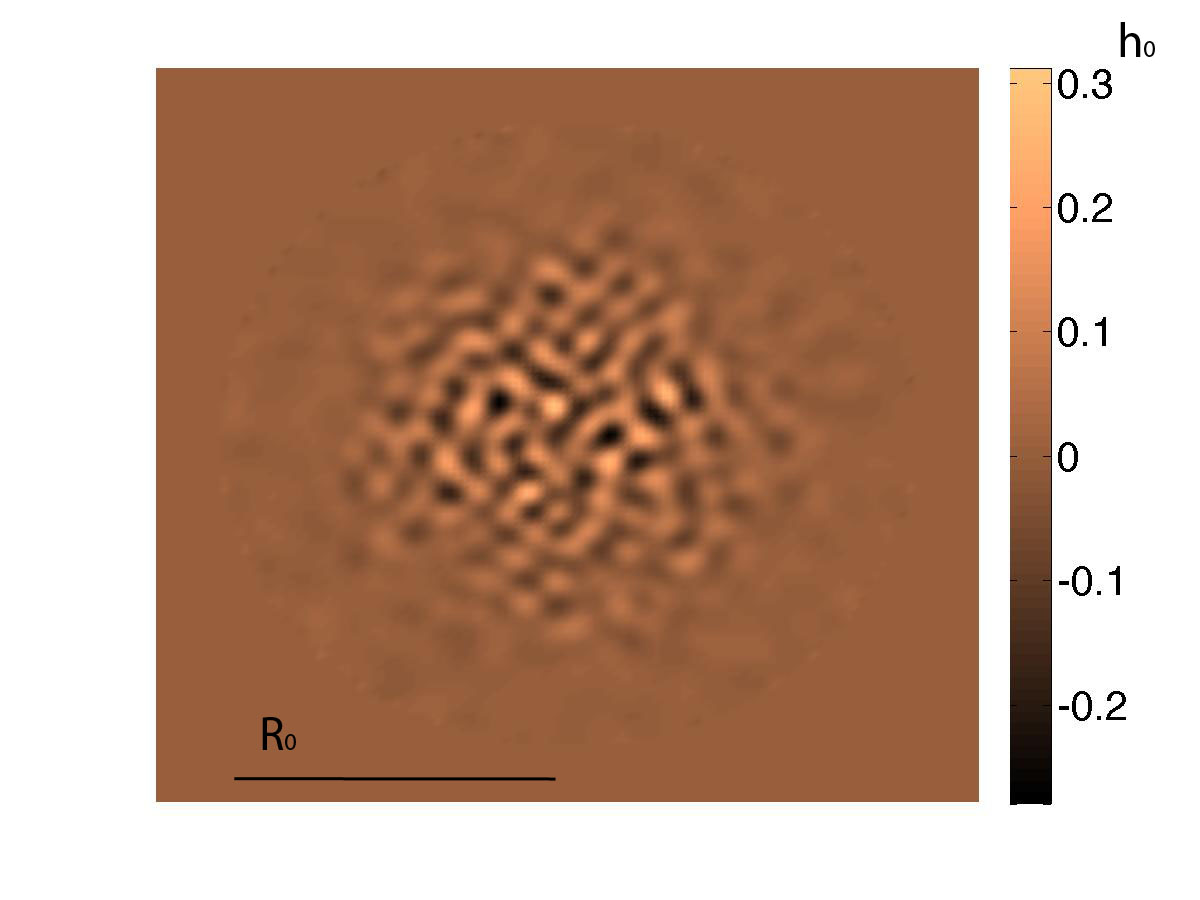}
\caption{Wrinkle formation and coarsening in a growing film
with residual stresses computed from analytical formulas for the 
pressures. {As the approximation breaks down,
the height of the central wrinkles increases much faster than
the height of the outer ones, which blur in comparison.}
Snapshots taken at times (\textbf{a}) $1.8/g$, (\textbf{b}) $2/g$, (\textbf{c}) $2.2/g$, 
and (\textbf{d}) $2.4/g$, starting from a randomly perturbed biofilm of radius 
$R_0=10^{-3}$ m and height $h_0=10^{-4}$ m. The radius does
not vary significantly during this time, whereas the height becomes of
the order of the radius at the end. Parameter values: 
$1/g=2.3$ hours, $\mu_f=8.9 \times 10^{-4} $ Pa$\cdot$s at $25^o$, 
$\xi_\infty=70$ nm, $\phi_\infty=0.2$,  $h_a=100 h_0$, $\Pi=30$ Pa 
(taken from \cite{seminara}), $E=25$ kPa (taken from \cite{asally}), 
$\nu=0.4$, $\mu_s=  8.92$ kPa, $\nu_a=0.45$, $\eta_a=1$ kPa$\cdot$s, 
$\mu_a=0$, $h_{inf}=h_0/10$.}
\label{fig4}
\end{figure}

The residual stresses $\boldsymbol \sigma^0$ in (\ref{stress}) can be estimated 
averaging the osmotic and fluid pressure contributions to the three-dimensional 
biofilm. If the solution of (\ref{finalphis})--(\ref{finalvs}) in the biofilm $\Omega_b(t)$ 
is known, $\sigma_{ij}^0 $ could be estimated from
\begin{eqnarray}
-\int_{x_3=\xi}^{x_3=h} [(p_f+\pi_f) \mathbf I] \, dx_3,
\quad i,j=1,2,
\label{residualstress}
\end{eqnarray}
where $x_3=h$ and $x_3=\xi$ define the two biofilm interfaces with air and agar, 
see Figure \ref{fig3}.  Analytical approximations (\ref{echeight})--(\ref{selfsimilarhR})
of the the biofilm height $h$ in early stages of the biofilm evolution allow for 
simple simulations of the onset of wrinkle formation. Figure \ref{fig4} uses 
these asymptotic profiles to compute pressures and velocities  by means of
(\ref{p})--(\ref{v}). Starting from an initially flat biofilm, (\ref{residualstress}) 
suggests that we should consider stress profiles of the form
$-A h^2/2 - A R h - \Pi \phi_{\infty} $,
where $A={g\mu_f \over \xi_{\infty}^2 (1-\phi_{\infty})^2},$ which are 
nondimensionalized dividing by $E$. The first two terms reflect the 
stresses due to growing height and radius, whereas the last one accounts 
for the osmotic pressure.
Inserting these residual stresses in (\ref{plategrowth1bis})--(\ref{stress})
we generate small  inhomogeneities and wrinkles  in Figure \ref{fig4}.
However, these approximations neglect spatial variations in concentrations, 
as well as changes in cell behavior, and therefore, in stresses and
pressures. Therefore, the patterns display soon an unrealistic behavior,
with wrinkles excessively growing in the central region.

Solving the full set of coupled equations we have derived is very 
costly and  faces severe numerical difficulties at contact points.
Alternatively, we may set $\boldsymbol \sigma^0=0$ and work with the 
residual strains $\boldsymbol \varepsilon^0$ in~(\ref{strain}), which can 
be related to growth tensors created by stochastic cell processes as we
discuss next.

\section{Incorporating Cellular Behavior}
\label{sec:stochastic}

Cells within a biofilm differentiate to perform different tasks, and can 
deactivate due to lack of resources or die as a result of biochemical 
stress and waste accumulation \cite{asally,hera}. 
Such variations in the biofilm microstructure affect the overall shape
\cite{espeso}. Cell activity enters the previous deterministic model
through the biomass creation term $g(c_n) \phi_s$ in (\ref{finalphis}),
the nutrient consumption term $r_n(\phi_s,c_n)$ in~(\ref{lcnutrient}), and
the residual stresses $\boldsymbol \sigma^0$ in (\ref{stress}).
However, this does not account for cell death, cell deactivation and cell
differentiation. 

\begin{figure}[!hbt]
\centering
\includegraphics[width=12cm,angle=0]{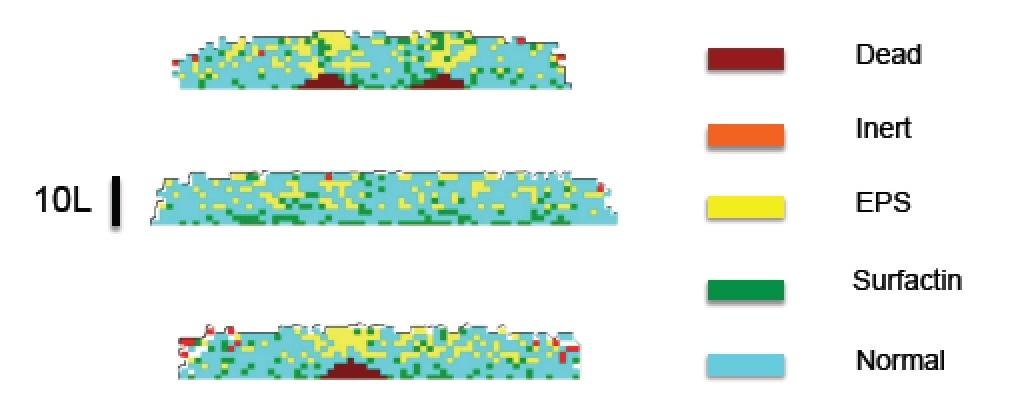}  
\caption{Layered distribution of dead, active, and inert cells, as illustrated
by slices of a growing biofilm. Dead cells appear at the bottom of three
peaks present in the initial biofilm seed.}
\label{fig6}
\end{figure}

Differentiation implies changes in phenotype while preserving the same 
genotype. For {\em B. subtilis} biofilms, the differentiation chain through 
which different cell types originate is established in \cite{hera},
see Figure \ref{fig6}. 
Initially, we have a population of similar alive cells glued together
in a matrix, most of which have lost their individual motility. All of them 
secrete ComX. If the concentration of ComX becomes large enough, 
some cells differentiate and start producing surfactin, losing their
ability to reproduce. For large enough surfactin concentrations,
other normal cells differentiate and become EPS producers.
These cells reproduce more slowly than normal cells. Cells may also 
die due to biochemical stresses \cite{asally}, preferentially at high-density 
regions,  such as the  agar--biofilm interface. In the upper
regions of the biofilm, depletion of resources may trigger deactivation
of cells, which become spores. Undifferentiated
cells retaining some motility are restricted to the bottom edges \cite{hera}. 

To a large extent, these processes have a random character. Hybrid 
models combine stochastic descriptions of cellular processes with continuous 
equations for other relevant fields. This allows us to consider the inherent 
randomness of individual bacterial behaviors as well as local variations 
\cite{hera,wingreen2}. We will explain how to introduce cell variability in
the mixture model next.



\subsection{Cellular Automata and Dynamic Energy Budget}
\label{sec:adeb}

In a cellular automata approach, space is divided in a grid of
cubic tiles. This grid is used to discretize all the continuous
equations: concentrations, deformations, pressures, etc. To simplify
geometrical considerations, we initially assume that each tile can
contain one bacterium at most. We describe bacterial metabolism
using a dynamic energy budget  framework \cite{birnir,deb}. 
According to this theory \cite{deb}, cells create energy from 
nutrients/oxygen, which they use for growth, maintenance, and 
product synthesis. Damage-inducing compounds can cause death.
The metabolism of each cell ${\cal C}_j$ is described by the system:
\begin{eqnarray}  
\begin{array}{ll}
{de_j \over dt} =   \nu' (f-e_j),  &f ={c_n \over c_n +K_n},  \quad
\nu' = \nu e^{- \gamma \varepsilon},  
\\
{dv_j\over dt} =  \Big(r_j {a_j \over a_M} -h_j \Big) v_j,   &
 r_j = \Big({\nu' e_j - m  g \over e_j+g} \Big)^+,   \\
{dv_{e,j} \over dt} = (1-\alpha) r_{e,j} v_j, & 
 r_{e,j} = k r_j + k',   \\
{dq_j \over dt}  =  e_j(s_G \rho v_j \,q_j \!+\! h_a) (\nu' \!- \! r_j)  
- (r_j \!+\! r_{e,j}) q_j,
&  \\  
{dh_j \over dt}  =  q_j - (r_j+r_{e,j}) h_j,  & \\  
{dp_j \over dt}  =  - h_j p_j, & p_j(0)=1,    \\  
{da_j \over dt}  =  (r_j+r_{e,j}) ( 1- {a_j \over a_M}),& \\
\end{array} \label{debsystem}
\end{eqnarray}
where the cell variables are  their energy $e_j(t)$,  volume $v_j(t)$, 
produced volumes of EPS $v_{e,j}(t)$,   
acclimation $a_j(t)$,  damage $q_j(t)$,  hazard $h_j(t)$, and 
survival probability $p_j(t)$, for $j=1,\ldots,N$, $N$ being the total 
number of cells. These equations are informed by the value
of continuous fields at the cell location $\mathbf x$ (the tile of the cellular
automata grid containing the cell):  nutrient concentration $c_n(\mathbf x,t)$, 
polymeric matrix concentration $c_e(\mathbf x,t)$, surfactin concentration 
$c_s(\mathbf x,t)$, ComX concentration $c_{cx}(\mathbf x,t)$, and environmental 
degradation $\varepsilon(\mathbf x,t)$, which are governed by
\begin{eqnarray}   
\begin{array}{ll}
{d c_n \over dt}  = - \nu' f \rho \sum_j v_j \delta_j + {\rm div}(d_n \nabla c_n) 
- {\rm div}(\mathbf v_f c_n),  \\[1ex]
{d c_{cx} \over dt}  = \rho \sum_j r_{cx,j} v_j \delta_j + {\rm div}(d_{cx} \nabla c_{cx})
- {\rm div}(\mathbf v_f c_{cx}), \\[1ex]
{d c_{s} \over dt}  =  \rho \sum_j r_{s,j} v_j \delta_j + {\rm div}(d_s \nabla c_{s})
- {\rm div}(\mathbf v_f c_s), \\[1ex] 
{d c_{e} \over dt}  =  \alpha \rho \sum_j r_{e,j} v_j \delta_j + {\rm div}(d_e \nabla c_{e})
- {\rm div}(\mathbf v_f c_e),  \\[1ex]
{d \varepsilon \over dt}  = \nu_{\varepsilon} \sum_j (r_j + \nu_m m) v_j  \delta_j 
+ {\rm div}(d_\varepsilon \nabla \varepsilon) -  {\rm div}(\mathbf v_f \varepsilon). \\
\end{array} \label{debconcentration}
\end{eqnarray}

The parameters $\nu$, $m$, $g$, $a_M$, and $\rho$ are the energy conductance, 
the maintenance rate, the investment ratio, the target acclimation energy, and mass 
density for the bacteria under consideration, respectively. Other coefficients are the 
multiplicative stress coefficient $s_G$, the maintenance respiratory coefficient $\nu_m$, 
the noncompetitive inhibition coefficient $K_v$, and the environmental degradation 
coefficients $\gamma$ and $\nu_\varepsilon$. 
The parameters $d_n$, $d_{cx}$, $d_s$, $d_e$, and $d_\varepsilon$ stand for diffusivities.
The Dirac masses $\delta_j$ are equal to $1$ at the location of cell ${\cal C}_j$ and
zero outside.

The production rates $r_{s,j}$ and $r_{e,j}$ are zero, except when the cell is a surfactin
producer, or an EPS producer, respectively. In the latter case, $r_{e,j}= k r_j + k'$,
where $k$ and $k'$ correspond to constants controlling the chemical balances for 
polymer production. The parameter $\alpha \in [0,1]$ regulates the fraction of
produced polymer that remains in a monomeric state and diffuses as $c_e$, instead
of becoming part of larger chains that remain attached to the cells forming the matrix
$v_e$. 

In this framework, bacteria ${\cal C}_j$ die with probability $1-p_j$. 
Taking a cellular  automata view, we modify the cell nature according to selected probabilities, which are defined in terms of concentration values at the cell location.
A normal bacterium becomes a surfactin producer with probability $p_{cx}={c_{cx} \over c_{cx} + K_{cx}^*}$ and an $EPS$ producer with probability $p_e={c_s\over K_s^* 
+c_s} \Big(1-{c_n\over K_n^* +c_n}\Big)$. Cells become become inert with probability
$p_i = 1- {c_n\over K_n^* +c_n}$. A non-surfactin-producer whose volume has 
surpassed a critical volume for division, divides with probability $p_d = {c_n\over K_n^* +c_n}$.  Figure \ref{fig6} represents the cell type distribution for a growing biofilm.
The simulation started from a circular seed with a diameter of $60$ cells, and
nonuniform height. Each colored box in the slices represents one cell. The brown 
areas representing dead cells appear at the bottom of three initial peaks. 
We set ${k_n L^2 \over d_n K_n}=0.01$, ${k_{cx} L^2 \over d_{cx} K_{cx}}=0.01,$ and
${k_{s} L^2 \over d_{s} K_{s} }=0.8.$, where $L$ is a reference length representing
the tile size (approximately the bacterium size $2$ $\mu$m).

\begin{figure}[!hbt]
\centering
\includegraphics[width=12cm,angle=0]{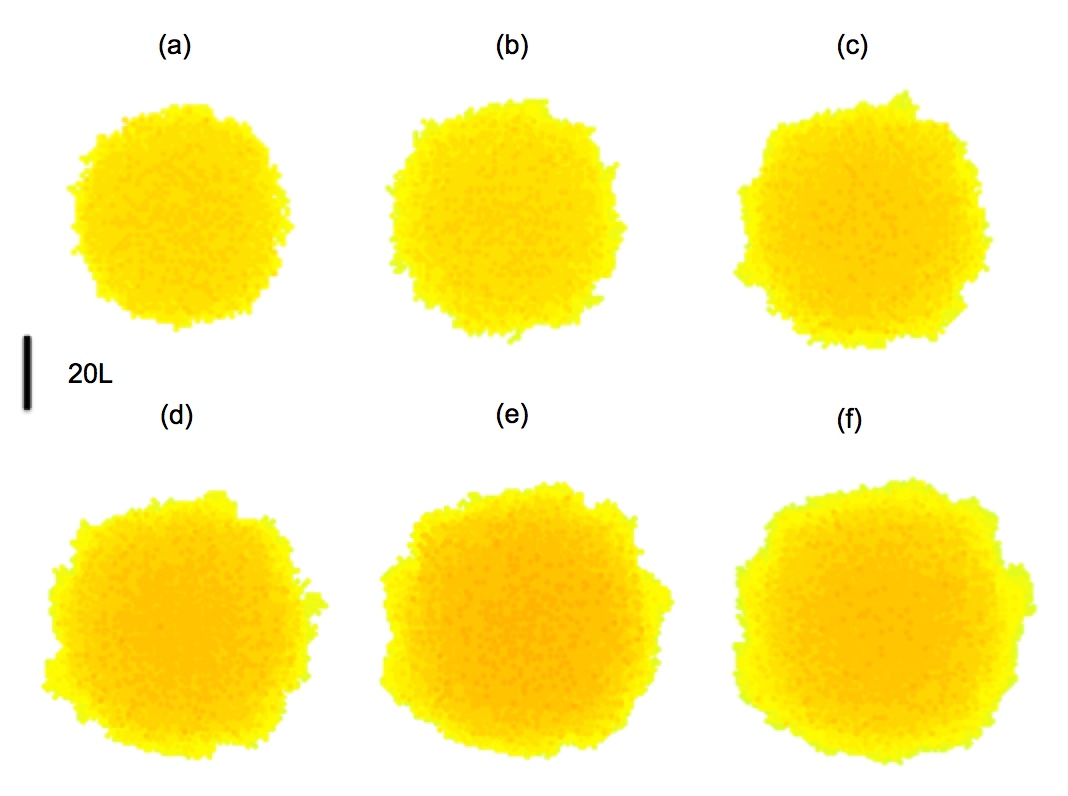}  
\caption{Top view of the evolution of a biofilm seed formed by two layers
of cells with a diameter of $40$ cells, {depicted at steps: ({\bf a}) $45$; ({\bf b}) 
$90$; ({\bf c}) $135$; ({\bf d}) $180$;  ({\bf e}) $225$;  ({\bf f}) $270$. } 
Darkest colors correspond to layers of increasing height up to $10$ cells. Contour undulations develop as the biofilm spreads.
}
\label{fig7}
\end{figure}

In principle, when a bacterium divides, the daughters occupy the space left by 
it, while pushing the other bacteria. Dealing with the geometrical aspects of 
arrangements of dividing bacteria is a complicate issue for which different
approaches have been explored \cite{ib1,ib2}; it is out of the scope of the
present work. For simplicity, we consider here that space is partitioned in a grid
of cubic tiles, as explained earlier, and this grid is used to discretize all the continuous 
equations for concentrations, displacements, pressures, etc.
Each tile may contain at maximum one bacterium, the tile size is the maximum 
size a bacterium may attain.
Once a bacterium divides, one daughter remains in the tile, whereas the other
occupies a random neighboring tile, either empty, or containing a dead cell,
which is reabsorbed. In the absence of them, it will push neighboring cells in
the direction of minimal mechanical resistance, that is, minimal distance to air.
The resulting collection of tiles defines the new $\Omega_b$. Figure \ref{fig7}
represents the evolution of a growing biofilm seed considering only
division processes with $c_n(0)=K_n$ and ${k_n L^2 \over d_n K_n}=8$.
Notice the development of contour undulations.

%

The resulting full computational model would proceed as follows. 

\vspace{6pt}
\noindent{\it Initialization}.

\begin{itemize}
\item We set an initial distribution of $N$ bacteria characterized by their
energies $e_j(0)$, volumes $v_j(0)$, damage $q_j(0)$, hazard
$h_j(0)$, acclimation $a_j(0)$, and attached EPS volume $v_{e,j}(0)$,
$j=1,\ldots,N$.
\item Each bacterium is initially classified as normal, surfactin producer, 
EPS producer, or inert.  Bacteria are distributed in the tiles $\mathbf x$
of the grid. The empty space around them is filled with water and dissolved 
substances. In this way, we may compute the volume fractions 
of biomass $\phi_s(\mathbf x,0)$ and fluid $\phi_f(\mathbf x,0)$ in each 
tile $\mathbf x$, as well as the osmotic pressure $\Pi(\mathbf x,0)$.
The pressure $p(\mathbf x,0)$ is obtained from (\ref{finaldarcy}) with 
$\mathbf v_s=0$ and $\boldsymbol \sigma^0(\mathbf x,0)$ from 
(\ref{residualstress}).
\item  We compute stationary solutions of the Equation 
(\ref{debconcentration}) for $\mathbf v_f=0$ by a relaxation 
numerical scheme.
All except the equation for $c_n$ are solved using the grid 
defining $\Omega_b(0)$ with no flux boundary conditions. The equation 
for $c_n$ is solved in the biofilm--agar domain, that is, $\Omega_b(0) \cup 
\Omega_a(0)$, imposing continuity of concentrations and fluxes at the 
 agar--biofilm interface and no flux boundary conditions at the
air--biofilm  interface.
\end{itemize}
{\it Evolution with a time step $dt$: From time $t-dt$ to $t$}.
\begin{itemize}
\item We update $\xi(t)$ using (\ref{plategrowth1bis})--(\ref{residualstress}).
Keeping the grid tile fixed, we shift the contains of the tiles upwards of 
downwards to reflect the evolution of $\xi(t)$ when the deflections are
large enough.  
\item We update the cellular fields solving (\ref{debsystem}) with the
stationary concentration values for (\ref{debconcentration}). The~time 
derivatives in (\ref{debconcentration}) are small, so that time evolution 
is driven by the source terms reflecting cell activity and changes in the 
biofilm boundaries.
\item We update the  bacterial status checking whether normal
bacteria become surfactin or EPS producers, whether any of them 
deactivates or dies, and whether they divide, with the probabilities assigned 
to each situation. In case a bacterium divides, we reallocate the newborn~cell. 
\item In the resulting biofilm configuration $\Omega_b(t)$, we compute the 
volume fractions of biomass $\phi_s(\mathbf x,t)$ and fluid $\phi_f(\mathbf x,t)$ 
in each tile. This also provides the osmotic pressure $\pi_f(\mathbf x,t).$
The fluid pressure $p(\mathbf x,t)$ is obtained from (\ref{finaldarcy}), the
displacements $\mathbf u_s(\mathbf x,t)$ from (\ref{finaldarcy}),
and $\boldsymbol \sigma^0(\mathbf x,t)$ from (\ref{residualstress}).
The solid velocities are approximated by $\mathbf v_s(\mathbf x,t) 
\sim (\mathbf u_s(\mathbf x,t) - \mathbf u_s(\mathbf x,t-dt))/dt$. Then,
the fluid velocity is given by (\ref{darcy}).
\item We yet need to take into account water absorption from agar. 
To do so, we solve 
${\partial \phi_f \over \partial t} + {\rm div}(\mathbf v_f \phi_f) =0$
in the biofilm/agar system. Alternatively, we can solve only
in the biofilm, using $\phi_f=0$ at the biofilm/agar interface and 
${\partial \phi_f \over \partial \mathbf n} = {h\over R} \phi_f$
for boundary conditions, where $h$ and $R$ are reference
values for the biofilm height and radius. Then, we revise the
biofilm configuration, creating water tiles with probability $\phi_f$
and shifting the contains of the neighbouring tiles. This provides
the final biofilm configuration $\Omega_b(t)$, that is, the
occupied tiles, their contents, the bacterial status and fields,
as well as the values of the continuous fields at each tile.
\end{itemize}

\begin{figure}[!hbt]
\centering
(\textbf{a}) \hskip 4.5cm (\textbf{b}) \hskip 4.5cm (\textbf{c}) \\
\includegraphics[width=5cm,angle=0]{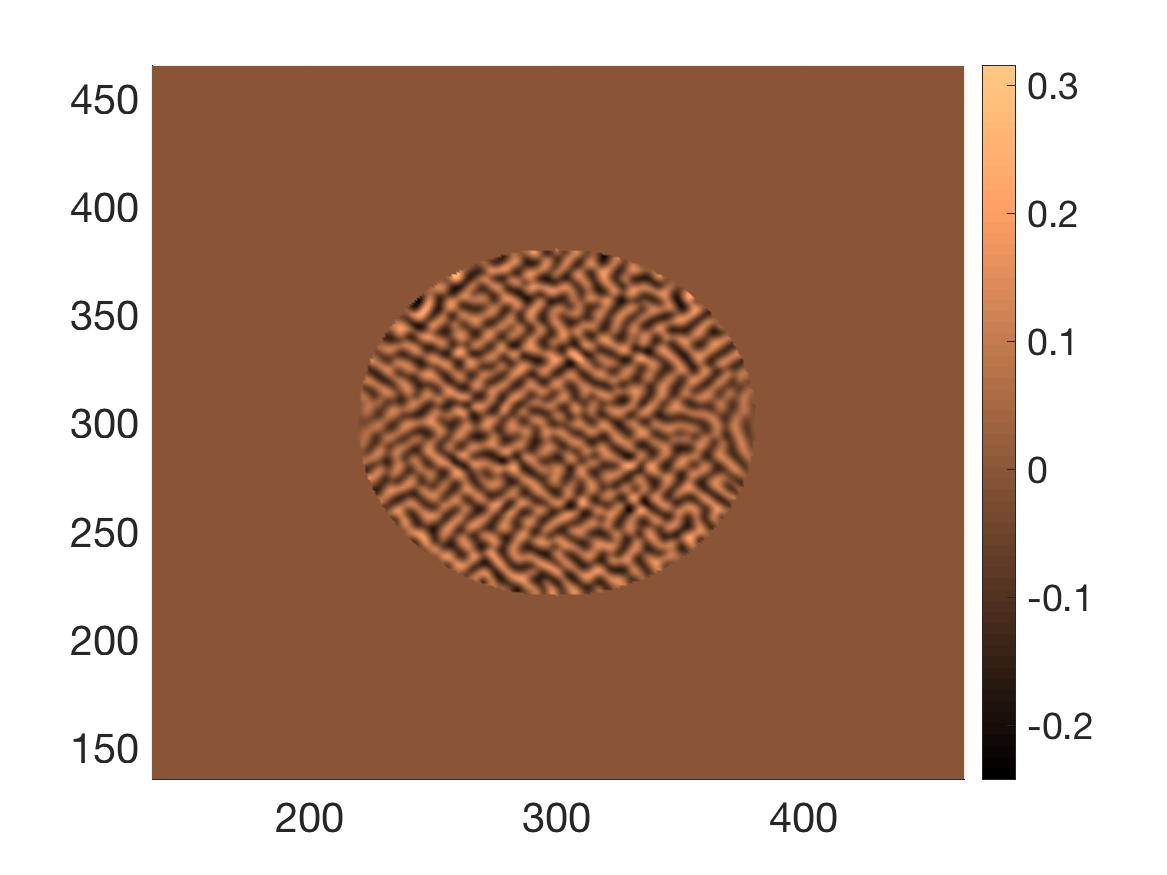}  
\includegraphics[width=5cm,angle=0]{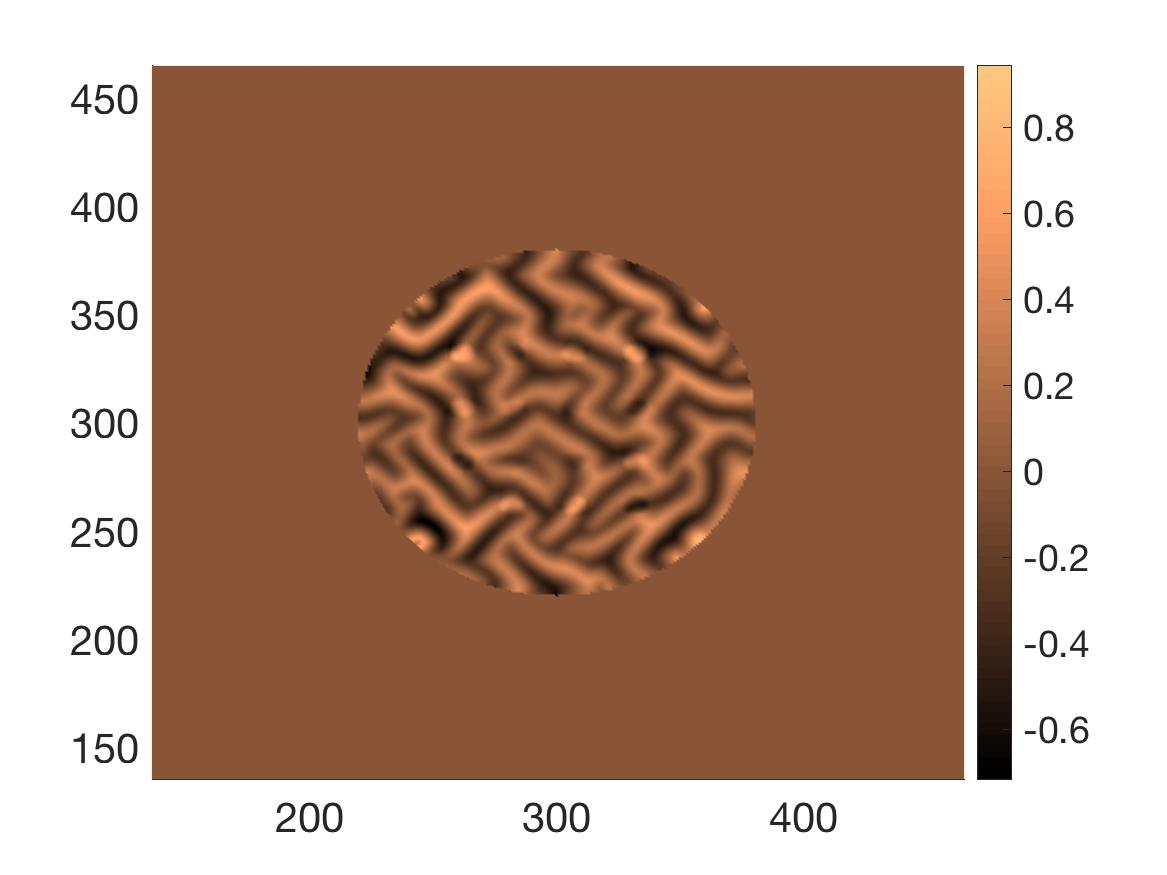}  
\includegraphics[width=5cm,angle=0]{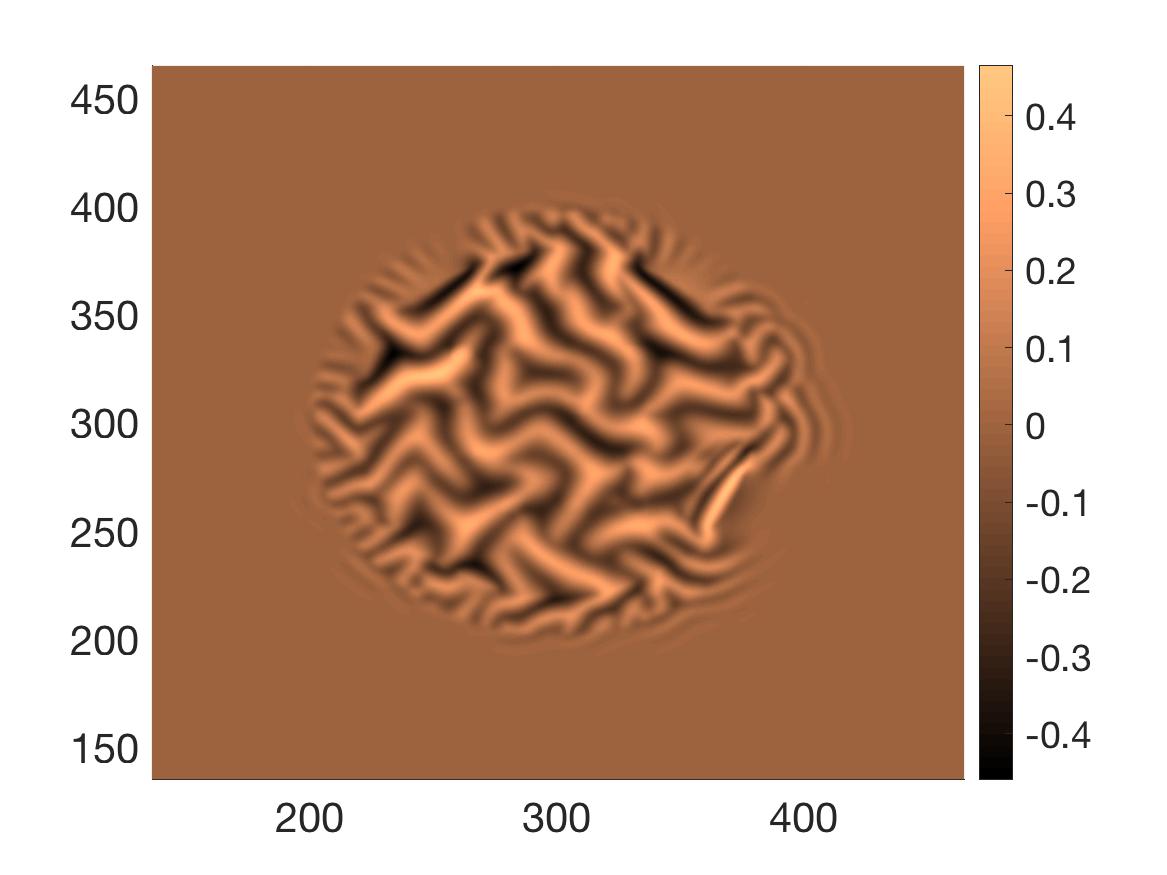} \\  
(\textbf{d}) \hskip 4.5cm (\textbf{e}) \hskip 4.5cm (\textbf{f}) \\
\includegraphics[width=5cm,angle=0]{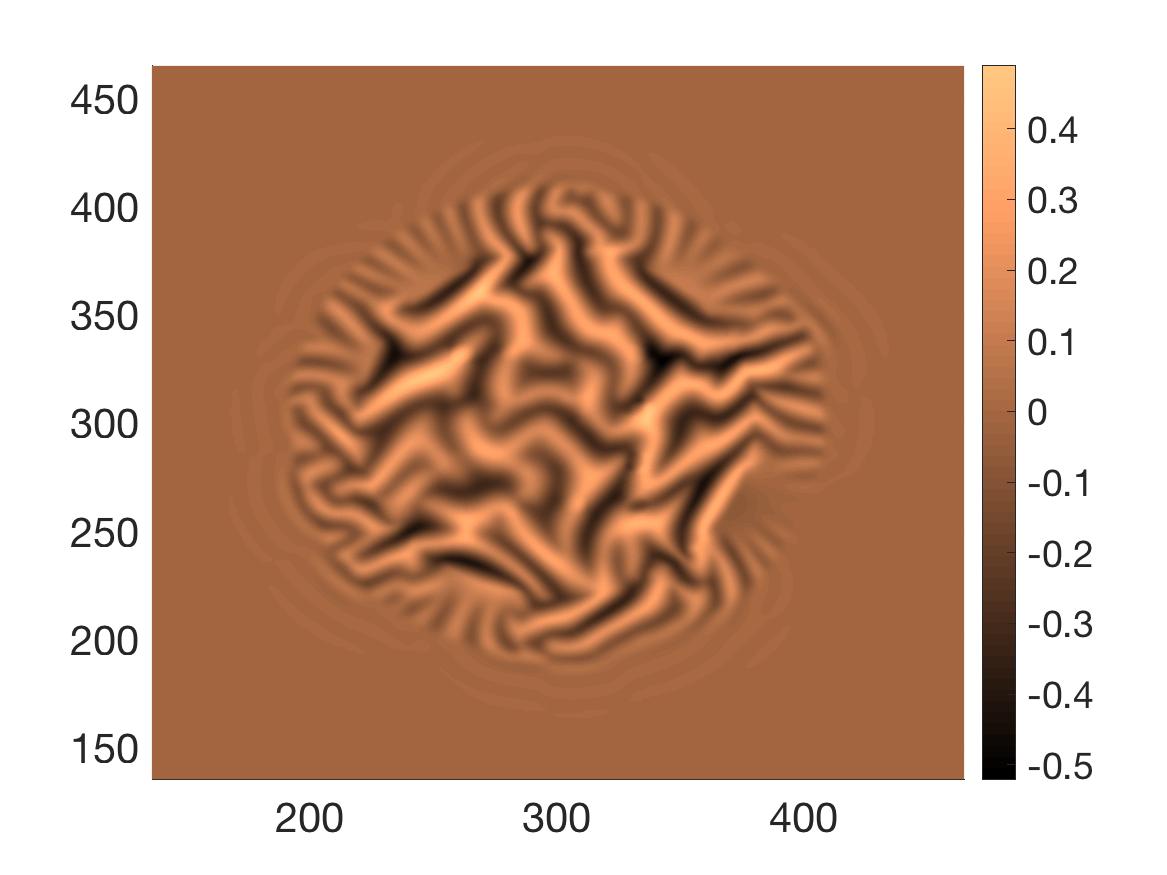}
\includegraphics[width=5cm,angle=0]{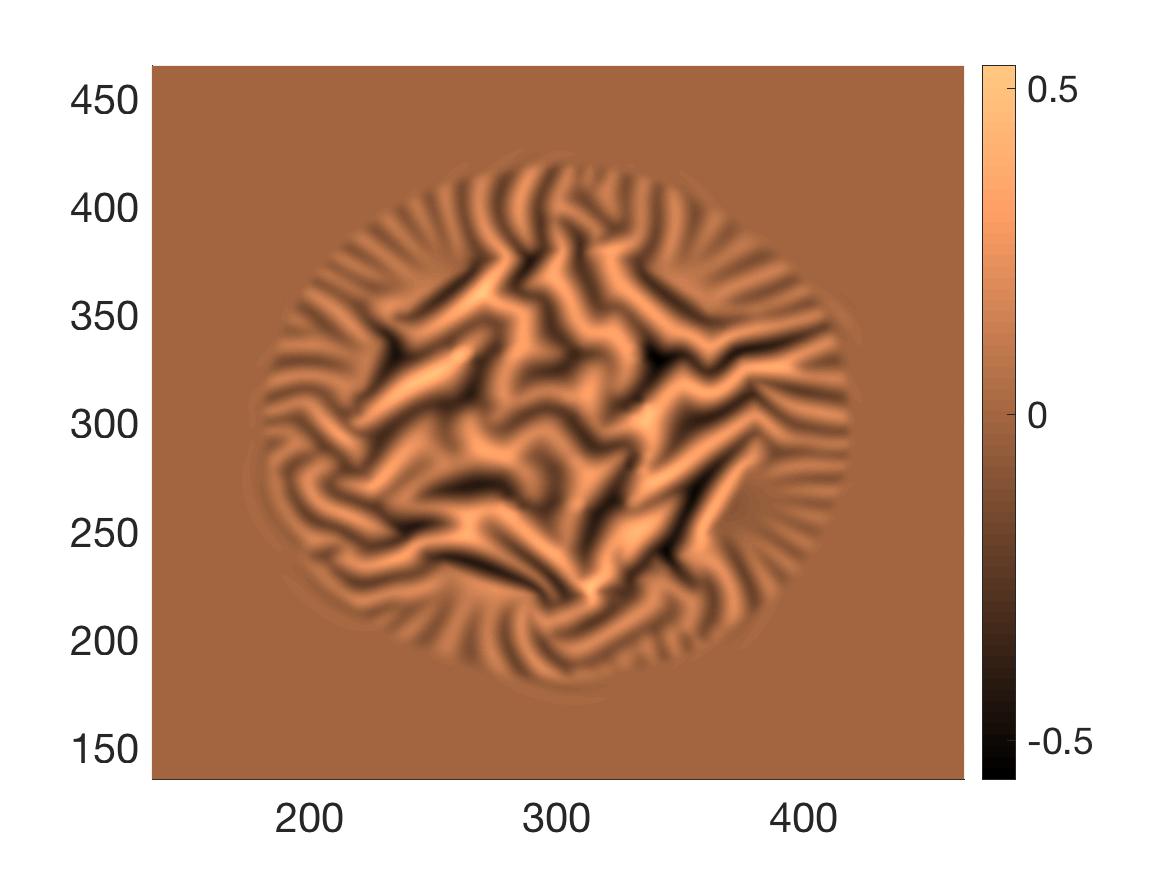}  
\includegraphics[width=5cm,angle=0]{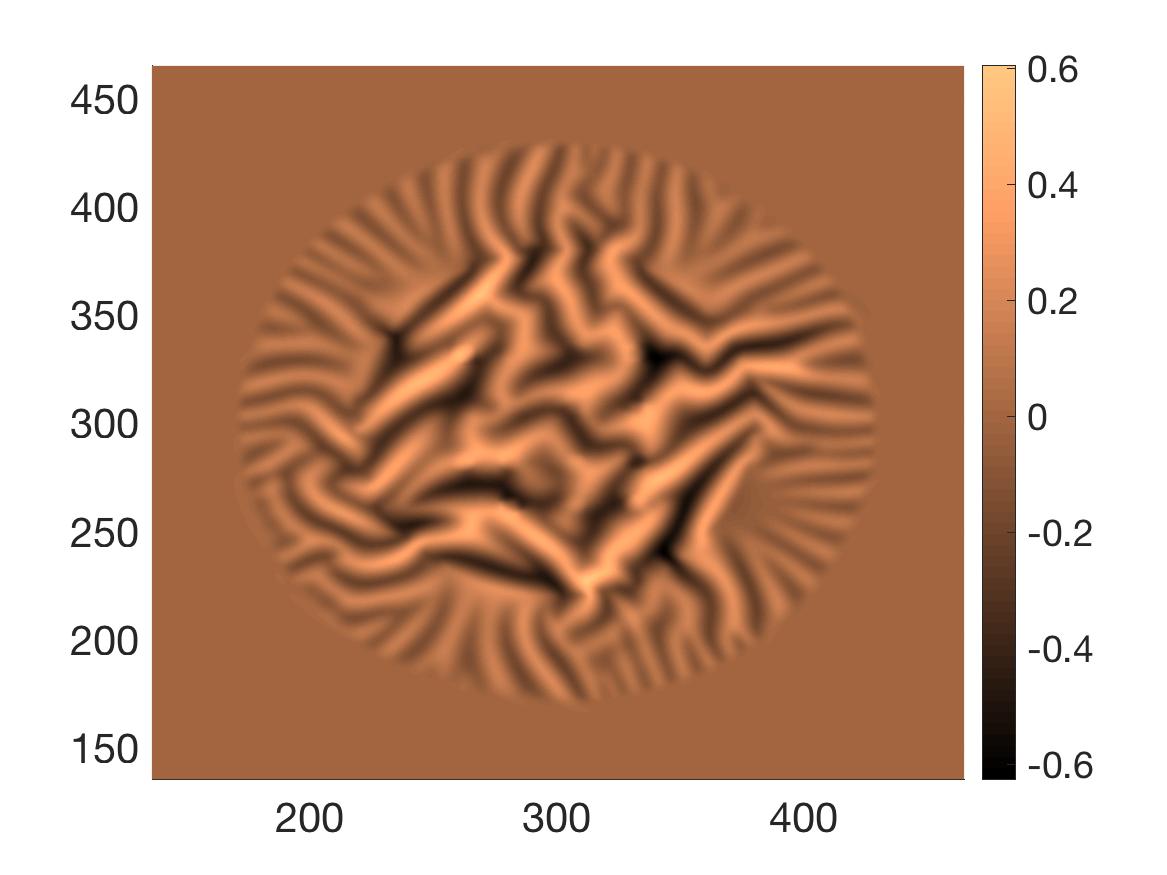} \\ 
(\textbf{g}) \hskip 4.5cm (\textbf{h})  \\
\includegraphics[width=5cm,angle=0]{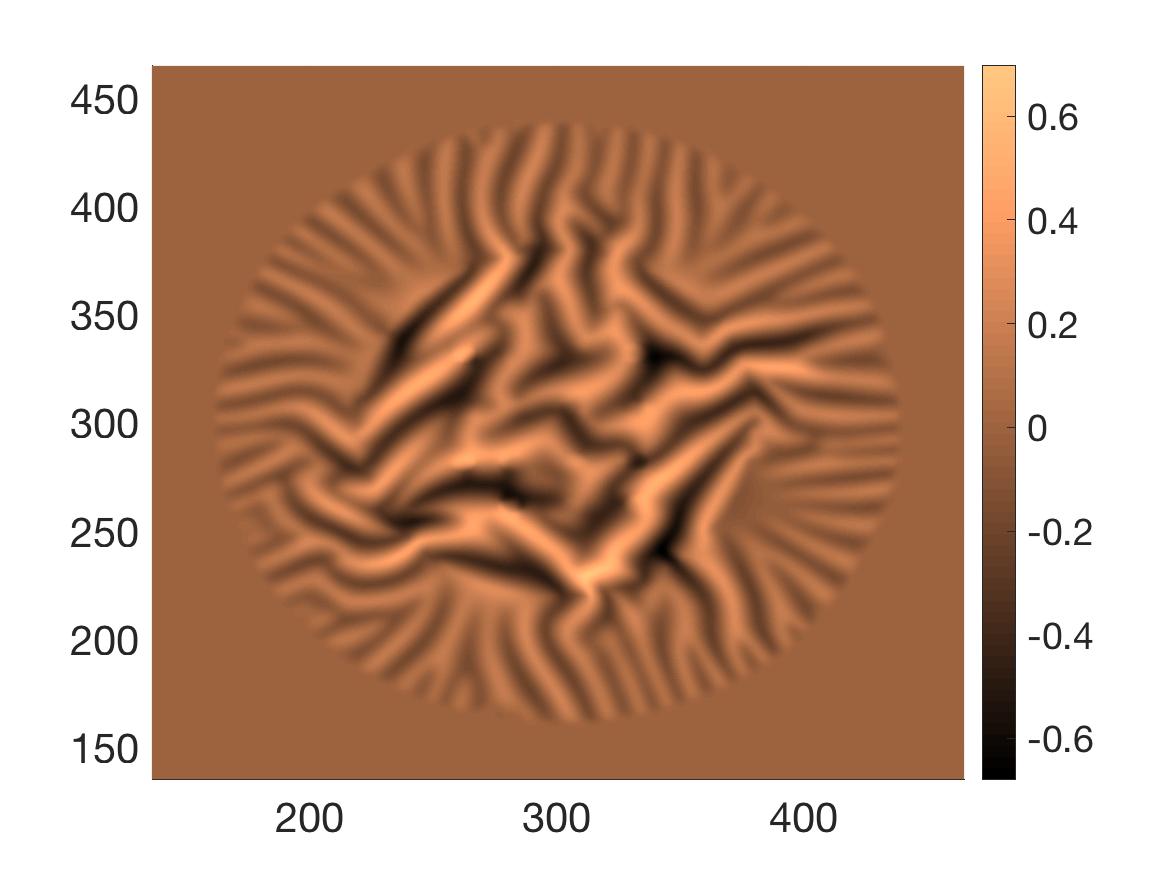}
\includegraphics[width=5cm,angle=0]{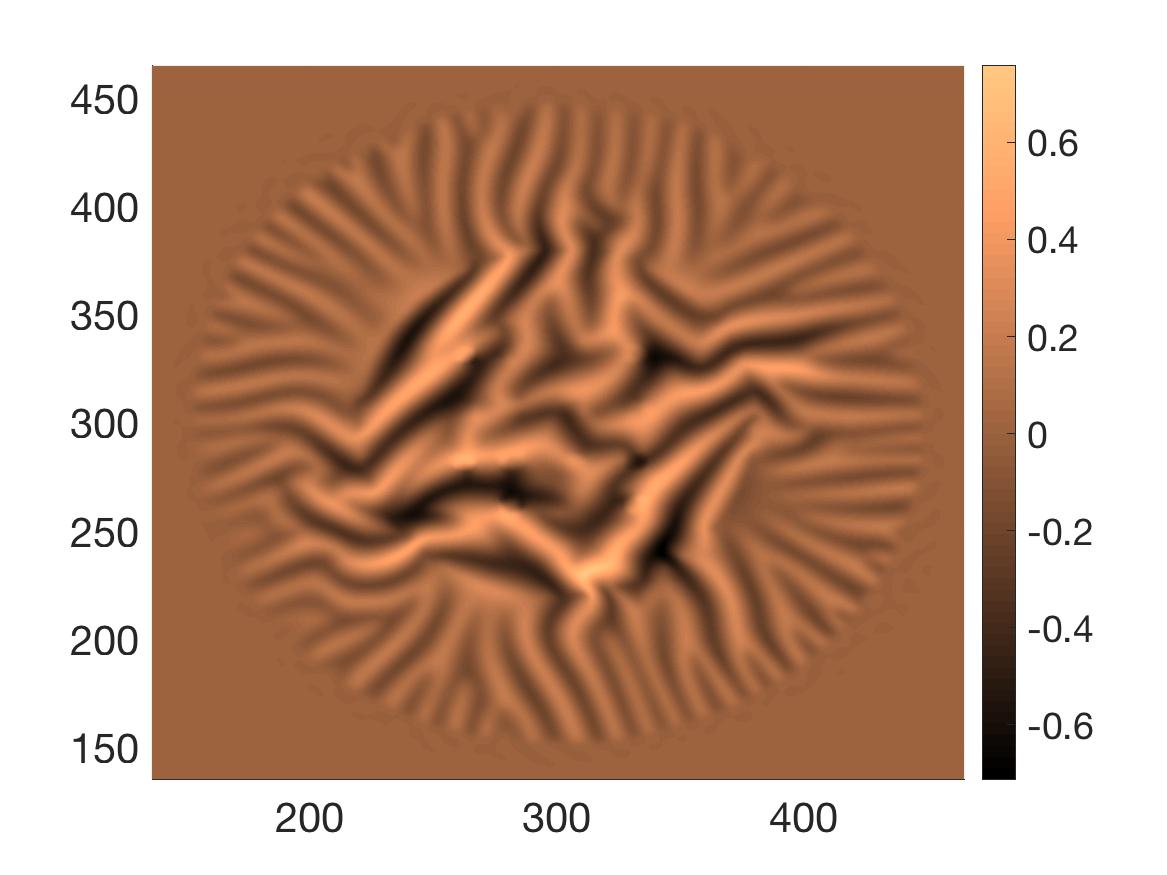} 
\caption{{Snapshots of} wrinkle formation, coarsening and branching as a circular
biofilm expands following (\ref{plategrowth1adim}) and  (\ref{plategrowth2adim})
and using an empirical fit to the residual stresses generated by cellular
processes. The biofilm has Poisson ratio $\nu = 0.5$ and Young
modulus $E = 25$ kPa. The Poisson ratio and rubbery modulus of the
substratum are $\nu_{v} = 0.45$, $\mu_v = 0,$ and $\gamma=16$. 
{ ({\bf a}) $26 {T\over \tau}$ s; ({\bf b}) $260 {T\over \tau}$ s; 
({\bf c}) $520 {T\over \tau}$ s; ({\bf d}) $780 {T\over \tau}$ s;
({\bf e}) $1040 {T\over \tau}$ s; ({\bf f}) $1300 {T\over \tau}$ s; 
({\bf g}) $1560 {T\over \tau}$ s; ({\bf h}) $1820 {T\over \tau}$ s;}
}
\label{fig8}
\end{figure}

This process mixes the stochastic evolution of some cellular
processes with continuous equations for a number of fields.
In case, any of the fields computed from the tile configuration
is not smooth enough to solve the required partial
differential equations, we filter it using a total variation based
filter introduced  in \cite{poroelastico} to avoid such artifacts.
Figure \ref{fig8} depicts the formation, coarsening, and  branching 
of  wrinkles in an spreading biofilm when the residual stresses
are fitted by an empirical circular front approximation of
magnitude $-0.1$ advancing one grid box every $14/\tau$
seconds.

This hybrid model also introduces a number of parameters that 
should be calibrated to experimental data, not yet available.
The parameters appearing in the dynamic energy budget equations 
have been fitted to experimental measurements for {\it 
Pseudomonas aeruginosa} biofilms under the action of antibiotics
\cite{birnir}; fitting to {\it Bacillus subtilis} would require new specific
experiments.

\subsection{Balance Equation Approach}
\label{sec:vf}

\begin{figure}[!hbt]
\centering
(\textbf{a}) \hskip 2.5cm  (\textbf{b})  \hskip 2.5cm  (\textbf{c})  \hskip 3cm\\
\includegraphics[width=3cm]{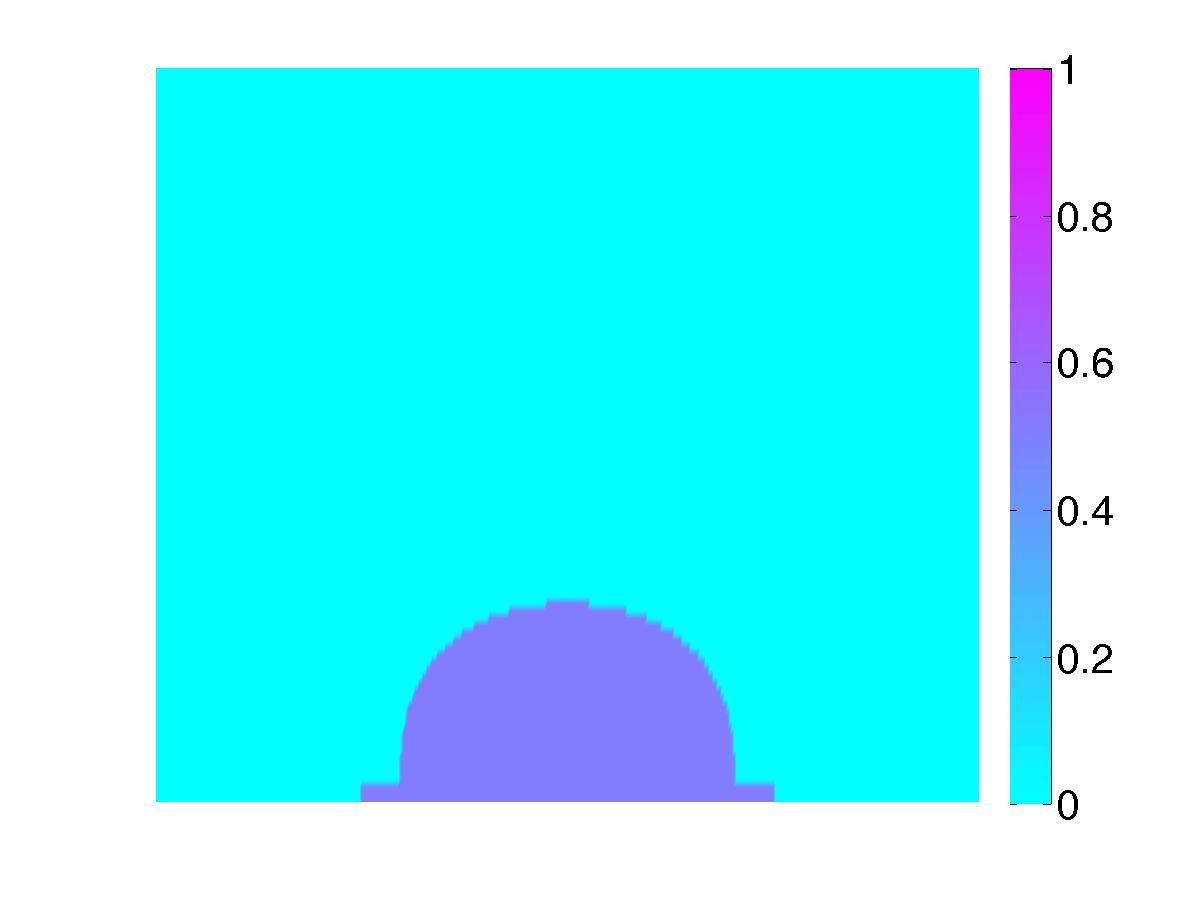}  
\includegraphics[width=3cm]{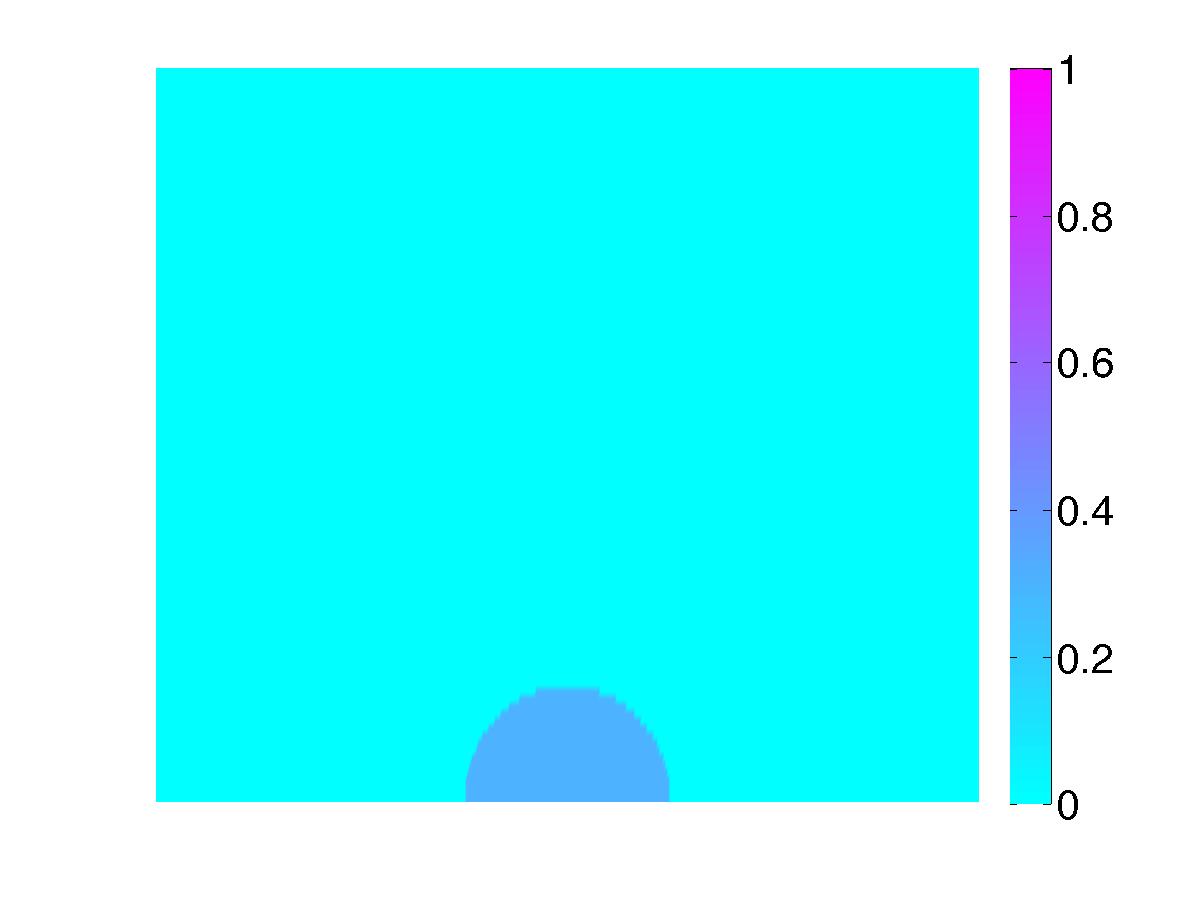}  
\includegraphics[width=3cm]{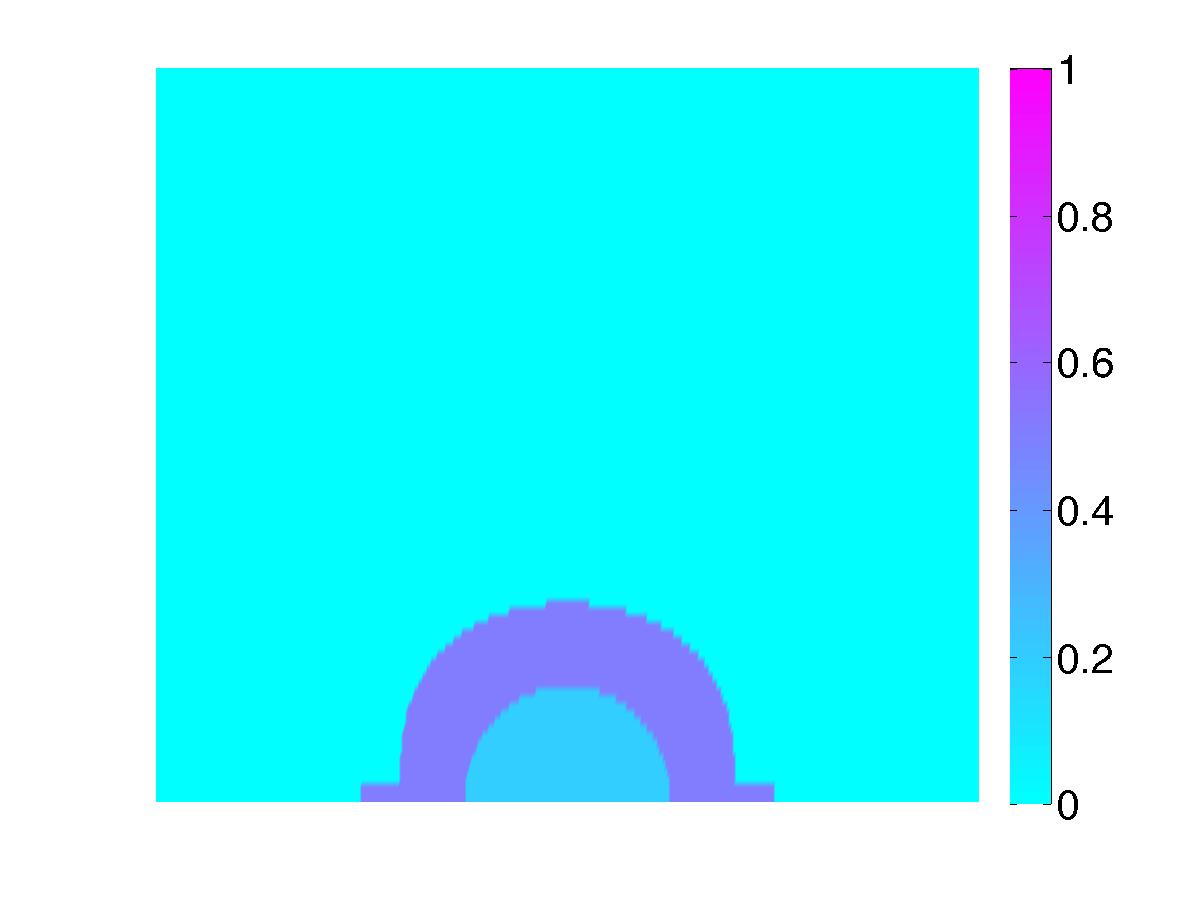}   \\
(\textbf{d})  \hskip 2.5cm  (\textbf{e})  \hskip 2.5cm  (\textbf{f})   \hskip 3cm\\
\includegraphics[width=3cm]{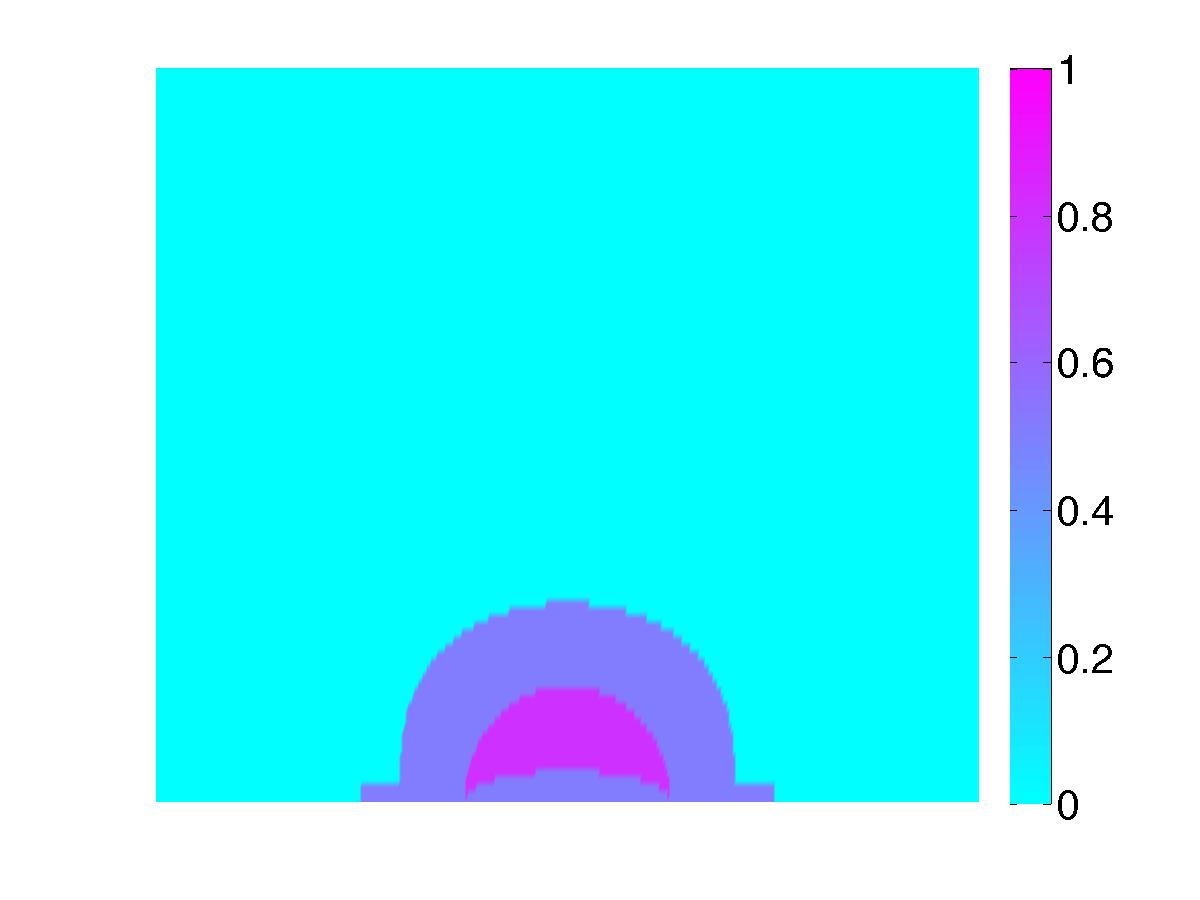}  
\includegraphics[width=3cm]{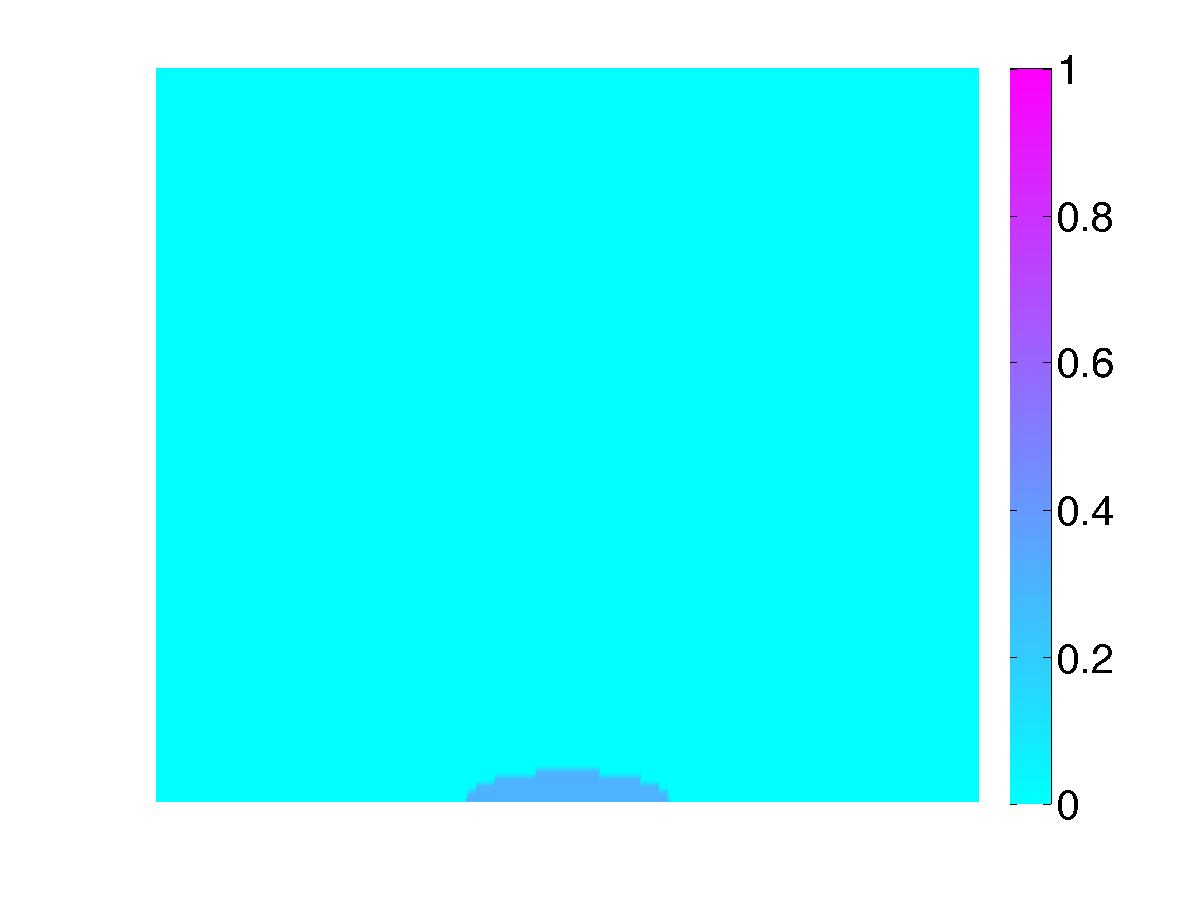}  
\includegraphics[width=3cm]{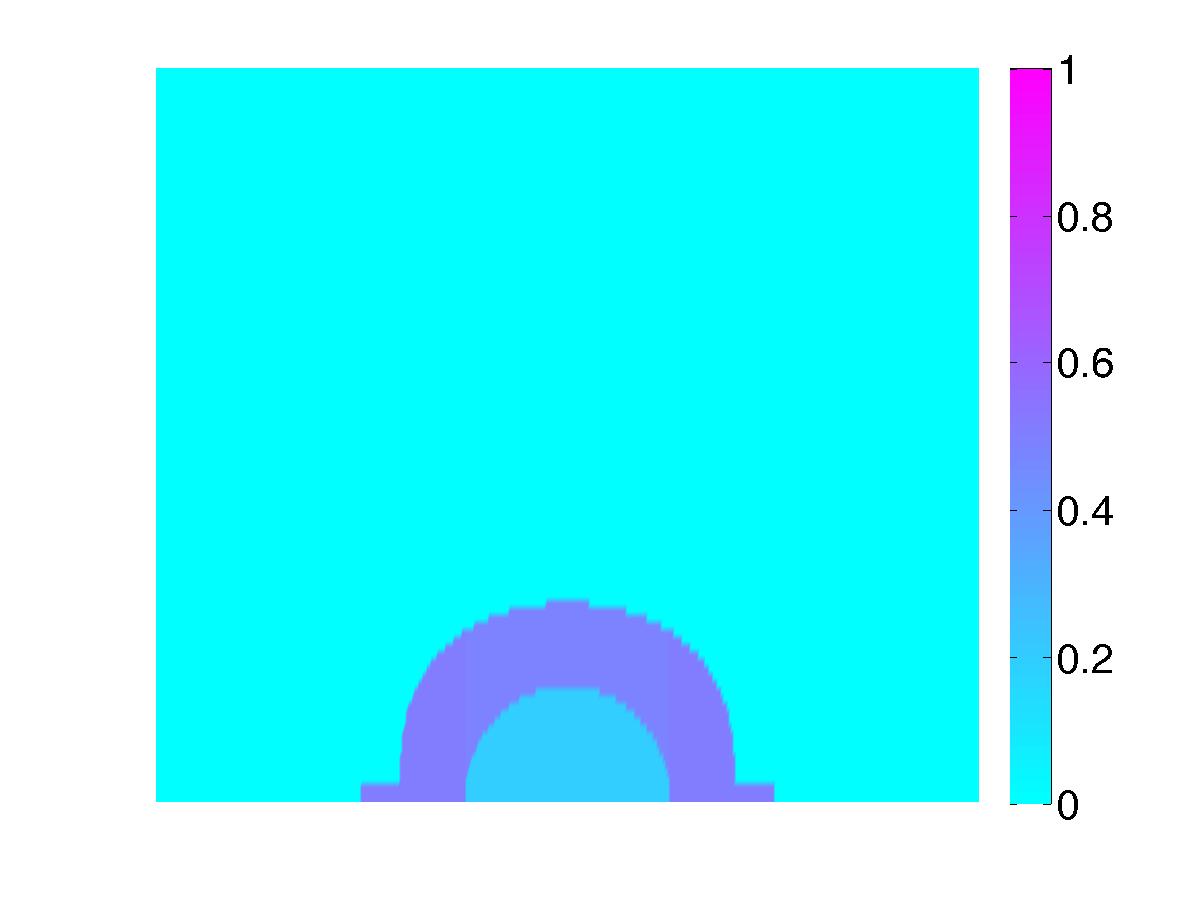}   \\
\caption{Effect of the presence of dead regions in liquid transport.
Initial volume fractions: (\textbf{a}) water, (\textbf{b}) dead cells, and (\textbf{c}) alive cells.
Snapshot showing dead cell reabsorption and water accumulation
in the originally dead regions at a later time: volume fractions of
(\textbf{d}) water, (\textbf{e}) dead cells,  and (\textbf{f}) alive~cells.}
\label{fig10}
\end{figure}

The macroscopic effect of the presence of differentiated bacteria
can partially be understood by means of additional balance equations,
inserting specific information in them.
Let us set $\phi_a$ and $\phi_d$  as the volume 
fraction of active and dead cells, respectively. We introduce an additional 
volume fraction of inert cells $\phi_i$, in such a way that
$\phi_s=\phi_a+\phi_i+\phi_d. $
The balance equations become
\begin{eqnarray}
{\partial \phi_a \over \partial t} + {\rm div}  (\phi_a {\mathbf v}_s) 
= [g(c_n) -g_w(c_w)- g_i(c_n)^+] \phi_a + g_i(c_n)^- \phi_i,  
\label{finalphiai} \\ [-1ex]  
{\partial \phi_d \over \partial t} + {\rm div}  (\phi_d {\mathbf v}_s) 
= g_w(c_w) \phi_a - k_r \phi_d,   \label{finalphidi} \\ [-1ex]  
{\partial \phi_i \over \partial t} + {\rm div}  (\phi_i {\mathbf v}_s) 
= g_i(c_n)^+ \phi_a - g_i(c_n)^- \phi_i,   \label{finalphii} \\ [-1ex]    
{\partial \phi_f \over \partial t} + {\rm div}  (\phi_f {\mathbf v}_f) 
= - r_s(\phi_a,\phi_d,c_n),   \label{finalphifi} 
\end{eqnarray} 
where $r_s(\phi_a,\phi_d,c_n)=g(c_n) \phi_a - k_r \phi_d$, 
$k_r$ being the rate of reabsorption of dead cells and $c_w$ the 
concentration of waste. The concentration of nutrients still obeys
(\ref{lcnutrient}), replacing $\phi_s$ by $\phi_a$ in the consumption
term, whereas the concentration of waste $c_w$ obeys a similar
reaction--diffusion equation with source $r_w(\phi_a)=k_w \phi_a$,
$k_w>0$. Here, 
$g_i(c_n)$ is positive for small enough values of $c_n$ and
negative otherwise. For instance, we might take
$g_i(c_n)= \alpha- {c_n \over c_n + K_n},$ $\alpha \in (0,1).$
We assume that dead and alive cells move with the velocity
of the solid biomass $\mathbf v_s$. Adding up Equations 
(\ref{finalphiai})--(\ref{finalphifi}), we recover the relations 
(\ref{balancemixture}) and (\ref{finaldarcy}).
The displacements of the solid biomass $\mathbf u_s$ still obey
(\ref{finalus}) with two modifications. 
First, the osmotic pressure $\pi_f$ depends only on the fraction 
of cells producing EPS, which must be alive. Thus, $\pi_f=\pi(\phi_a).$
Second, the elastic constants $\lambda_s$ and $\mu_s$ may 
vary spatially in case necrotic regions containing a large density of
dead cells or swollen regions appear.  We focus here on the effect 
of necrotic regions on liquid transport within the biofilm. 
Figure \ref{fig10} illustrates water accumulation in regions with an 
initially high volume fraction of dead cells. 

To investigate the spatial distribution of cells secreting
autoinducers, we introduce additional volume fractions
of active cells
$
\phi_a = \phi_{u} + \phi_{surf} + \phi_{eps},
$
where $\phi_{surf}$ and $\phi_{eps}$ stand for cells producing
surfactin and EPS respectively, whereas $\phi_{u}$ are undifferentiated
active cells. 
The balance equations governing the different subpopulations
are
\begin{eqnarray}
{\partial \phi_u \over \partial t} + {\rm div}  (\phi_u {\mathbf v}_s) 
 = g(c_n) \phi_u + g_{e}(c_n)\phi_{eps} 
 - [g_{c}(c_{cx})+g_{s}(c_{surf})] \phi_{u}   \label{finalphiu} \\  [-1ex]    
{\partial \phi_{surf} \over \partial t} + {\rm div}  (\phi_{surf} {\mathbf v}_s) 
 = g_{c}(c_{cx}) \phi_{u},  
 \label{finalphisurf} \\   [-1ex]  
{\partial \phi_{eps} \over \partial t} + {\rm div}  (\phi_{eps} {\mathbf v}_s) 
= g_{s}(c_{surf}) \phi_{u},
 \label{finalphieps} 
\end{eqnarray} 
where $g_{e}(c_n)= \beta g(c_n)$, $\beta\in (0,1)$, 
$g_{s}(c_{surf})= k_{surf}^* {c_{surf} \over c_{surf} + K_{surf}^*}$,
and $g_{c}(c_{cx})= k_{cx}^* {c_{cx} \over c_{cx} + K_{cx}^*}$.
The~autoinducer concentrations are governed by balance equations
of the form (\ref{lcnutrient}) with sources
$r_{surf}=k_{surf}(1- {c_{surf} \over c_{surf} + K_{surf} })$ and
$r_{cx}=k_{cx}(1- {c_{cx} \over c_{cx} + K_{cx} })$,
respectively, as well as no-flux boundary conditions at the
biofilm boundaries.

If we consider dead and inert cells too, systems 
(\ref{finalphiai})--(\ref{finalphifi}) should be updated replacing
$g(c_n) \phi_a$  by $g(c_n) \phi_u + g_e(c_n) \phi_{eps}$ in
Equation (\ref{finalphiai}) and in the definition of $r_s$ for
Equation (\ref{finalphifi}). Likewise, systems 
(\ref{finalphiu})--(\ref{finalphieps}) should be updated including
transfer to and from the inert status.

\section{Discussion}
\label{sec:discussion}

Growth of cellular aggregates involves mechanical, chemical,
and cellular processes acting in different time scales.  Bacterial
biofilms provide basic environments to test hypotheses and mathematical
models against experimental observations. Recent experimental
work with {\it Bacillus~subtilis} reveals a host of phenomena during
biofilm formation and spread. Different approaches have been exploited
to account for different aspects: thin film equations and two-phase flow  
models for accelerated spread caused by osmosis \cite{seminara}, 
elasticity theory for the onset of wrinkle formation~\cite{asally,sergei}, Von 
Karman-type approximations for wrinkle branching \cite{espeso,wrinkles}, 
and Neo Hookean models for contour undulations \cite{benamar1} and 
fold formation. In principle, poroelastic models allow to consider liquid 
transport and elastic deformation in a unified way \cite{poroelastico}, 
though detachment and blister formation require further developments 
\cite{wingreen}.
Current models take mainly a deterministic point of view, thus, random
cell behavior linked to fluctuations is poorly accounted for. However,
cell differentiation \cite{hera} to incorporate new phenotypes performing
new tasks, such as autoinducer and EPS matrix production, plays
a key role in biofilm development. Elementary cellular automata approaches
were  implemented in \cite{espeso,poroelastico} and used to
generate nonuniform residual stresses partially defining the biofilm shape.
Here, we develop a hybrid computational model, combining a solid---fluid
mixture description of mechanical and chemical processes with a
dynamic energy budget based cellular automata approach to cell 
metabolism. 

{Cellular automata representations are convenient from
a computational point of view, as they allow for simple rules to
transfer information between individual cells and the film. However,
they provide too crude a representation of bacterial geometry. In our
framework, this representation could be improved by resorting to different 
agent based models. Individual-based models, originally developed to study 
biofilms in flows \cite{ib0,ib3}, have recently been adapted to describe biofilms 
spreading over air--agar interfaces and solid--semisolid interfaces \cite{ib1,ib2}. 
Similarly, immersed boundary methods introduced to study bodies immersed 
in  fluids are being extended to study biofilm spread in flows~\cite{ibm_rheology} 
and at interfaces \cite{ibm_multicellular}. We could
resort to Individual based or Immersed boundary approaches for a better
description of bacterial geometry and their spatial arrangements.}

{Working with biofilms spreading on an air--agar interface,
we have chosen to represent the presence of small fractions of 
polymeric matrix in an effective way, as done in previous related work~\cite{seminara,basic1}.
The biomass formed by bacteria and polymeric threads is considered one 
phase \cite{seminara}, with elastic properties as in \cite{basic1}. The liquid
transporting dissolved chemicals is considered a fluid phase. Production of
EPS also affects internal liquid flow by osmosis, mechanism we include
in our equations for the fluid phase. Depending on the relative fractions 
and the properties of each phase as well as the characteristic times and 
lengths, the whole system may display an elastic, fluid, viscoelastic, or truly 
poroelastic behavior \cite{keller,kapellos}. 
This formulation allows to derive effective equations for the dynamics of the 
interfaces including the effect of biomass growth, fluid, and osmotic
pressures through residual strains and stresses. Resorting to individual-based 
or immersed boundary representations of cells, we might describe
the polymeric matrix as a network of threads instead \cite{ib1,ibm_rheology},
but we should define heuristic rules for their behavior.}

Constructing numerical solutions of the full model is 
a computational challenge, out of the scope of the present work.
Instead, we construct numerical solutions, in particular, geometries, 
guided often by asymptotic simplifications. 
In this way, we show that the model is able to reproduce behaviors 
experimentally observed: {
accelerated spread due to water intake \cite{seminara,wingreen}, 
wrinkle formation and branching \cite{asally,wilking,wingreen}, 
layered distributions of differentiated cells \cite{hera}, development
of undulations in the contour \cite{benamar1,wingreen}, and
appearance of regions containing a high volume fraction of water
\cite{wilking,wingreen}.  Existing models are devised to explain
specific behaviors in relation with particular experiments.
An advantage of our approach is that a single model can be used
to display all those behaviors and to simulate or even
analyze under which conditions  they are observed, as the 
model allows for asymptotic analysis in specific situations}.
The partial study of different phenomena also
suggests empirical expressions for magnitudes representing
cellular activities required by the mixture model, such as source
terms or residual stresses, which can be inserted in it to reduce
computational costs.
{Our~simulations of biofilm spread and wrinkle formation 
use parameter values experimentally measured for {\em Bacilus
subtilis} biofilms in \cite{seminara,asally}, producing reasonable 
qualitative and quantitative  results. 
However, the parameters for the dynamic energy budget
systems for cell metabolism, as well as those appearing in the
concentration equations are taken from {\em Pseudomonas
aeruginosa} studies \cite{birnir}. The probability laws for the
cellular automata model and the balance equations for
differentiated cell populations involve additional unknown 
parameters.} Thus, our model involves a collection of parameters
that should be fitted to experimental data, specially as far as
cell metabolism is concerned. Experimental measurements of
bacterial dynamics allowing to fit such parameters are yet missing.\\

{\bf Acknowledgements.} A.C. thanks R.E. Caflisch for hospitality during a sabbatical stay at the Courant Institute, NYU. This research has been partially supported by the FEDER/Ministerio de Ciencia, Innovaci\'on y Universidades -Agencia Estatal de Investigaci\'on  grant No. MTM2017-84446-C2-1-R (AC,EC) and the Ministerio de Ciencia, Innovaci\'on y Universidades ``Salvador de Madariaga'' grant PRX18/00112 (AC).

\section*{Appendix}
\label{sec:apA}

Here, we derive approximate expressions for the volume fractions,
velocity, and pressure fields, as well as equations for the height, by 
considering a simplified version of the equations presented in Section 
\ref{sec:mixture} for a slice in the plane $x_1x_3$, as in Figure 
\ref{fig3}a.

Only the components $x_1$ and $x_3$ of the variables are relevant.
To simplify the notation, in this section we take $\phi=\phi_s$,
$p_f=p$, $\pi_f=\pi = \Pi \phi$, $\mathbf u_s= \mathbf u$, 
$\mathbf v_s=\mathbf v$. The subsequent arguments follow those
in \cite{seminara} for biphasic fluid mixtures, with adequate modifications
when necessary.

Following the work in \cite{seminara}, we assume that $g$ is approximately constant
and $\phi_t  << g \phi $. Then, the governing equations become
\begin{eqnarray}
\begin{array}{l}
{\rm div} (\mathbf v \phi) = g \phi, \quad
{\rm div} (\mathbf v) = {\rm div} \left[ {(1-\phi)^2 \over \zeta}  
\nabla p \right], \\[2ex]
\mu_s \Delta \mathbf u + (\mu_s + \lambda_s) \nabla 
({\rm div}(\mathbf u)) =  \nabla (p + \Pi \phi),
\end{array} \label{equseminara}
\end{eqnarray}
where $\mathbf v= {\partial \mathbf u \over \partial t}.$
Recall that $\zeta= {\mu_f \over  \xi(\phi)^2}$, we will set
$\zeta = {\mu_f \over  \xi_\infty^2}$ with $\xi_\infty=\xi(\phi_\infty),$
$\phi_\infty$ being a background value.
We fix the displacements at the biofilm/agar interface $x_3=0$
and impose no stress conditions at the biofilm--air interface 
$x_3=h$, assuming that the normal vector $\mathbf n \sim (0,1)$:
\begin{eqnarray}
\begin{array}{l}
(u_1,u_3) \big|_{x_3=0} =(0,0), \quad
\sigma_{13} \big|_{x_3=h} = \mu_s (u_{1,x_3} + u_{3,x_1})\big|_{x_3=h} = 0, 
\\ 
\sigma_{33} \big|_{x_3=h} = \left[-( \pi_f + p_f ) + 2\mu_s u_{3,x_3} 
+ \lambda_s {\rm div} (\mathbf u) \right] \big|_{x_3=h} = -( \pi_f + p_f)_{ext}.
\end{array} \label{bcseminara}
\end{eqnarray}

Next, we nondimensionalize setting $u_i = U_i u_i'$, $v_i = V_i v_i'$, 
$x_1= R_0 x_1'$, $x_3=h x_3'$, $R=R_0 R'$, $p=Pp'$, with $h/R_0 << 1$. 
In practice, the height $h=h(x_1,t)$ depends on $x_1$ and $t$ and the
reference radius $R_0$ may vary with time.
If we  set $t= T t'$, with $T=1/g$, then $V_i=U_i g$. 
Changing variables and dropping the $'$ symbol to simplify, we find
\begin{eqnarray}
{V_3 \over g h} [\phi v_{3,x_3} +  \phi_{x_3}  v_{3} ] = \phi,  
{V_3 \over h} v_{3,x_3} = {P \over h^2 \zeta} [p_{x_3x_3} (1-\phi)^2  
- 2 (1-\phi) \phi_{x_3} p_{x_3}], \label{adim12}  \\
{\mu_s  U_1 R_0 \over P h^2 }u_{1,x_3x_3}  
 =  p_{x_1} +{\Pi \over P} \phi_{x_1},   
  (2 \mu_s + \lambda_s)  {U_3 \over Ph} u_{3,x_3x_3}
 =  p_{x_3} +  {\Pi \over P} \phi_{x_3}, \label{adim34} 
 \end{eqnarray}
with boundary conditions:
\begin{eqnarray}
\begin{array}{l}
(u_1,u_3) \big|_{x_3=0} =(0,0), \;
u_{1,x_3}\big|_{x_3=1} = 0,  \\
\left[-({\Pi \over P} \phi + p) + (2\mu_s+\lambda_s) 
{U_3 \over h P} u_{3,x_3} \right] \Big|_{x_3=1} =  
-({\Pi \over P} \phi + p)_{ext}. 
\end{array} \label{bcadim}
\end{eqnarray}

We can get an approximate solution of (\ref{adim12}) and (\ref{adim34})
in the same asymptotic limit as in 
\cite{seminara} with slight variations due to the fact that we have
an equation for the solid biomass displacements, 
not for fluid biomass velocities. 

First, set $V_3=gh$ and $P=gh^2\zeta$, to balance growth and flow in
(\ref{adim12}). For the rest, we  argue with 
(\ref{adim34}). We set  
$K_3= {\mu_s U_1 R_0 \over P h^2}$,
$K_2= {(2\mu_s + \lambda_s) U_3 \over P h} =
{2\mu_s + \lambda_s \over \mu_f} {\xi_\infty^2\over g h^2} $ 
and $\varepsilon_p={P\over \Pi}$, obtaining  
\begin{eqnarray} 
 (\phi_{x_3} v_3 + \phi v_{3,x_3}) = \phi,  & &
 v _{3,x_3} =  p_{x_3x_3} (1-\phi^2) - 2 p_{x_3} \phi_{x_3} (1-\phi), 
 \label{adim12bis} \\
 K_3 u_{1,x_3x_3}  
 =  p_{x_1} +  \varepsilon_p^{-1} \phi_{x_1},   & &
 K_2 u_{3,x_3x_3}  =  p_{x_3} +   \varepsilon_p^{-1} \phi_{x_3}, 
 \label{adim34bis}
 \end{eqnarray}
with 
$(u_1,u_3) \big|_{x_3=0} =(0,0),$ $ u_{1,x_3}  \big|_{x_3=1}= 0,$ $
\left[(- \varepsilon_p^{-1} \phi - p) + 2 K_2 u_{3,x_3} 
\right] \big|_{x_3=1} =  (- \varepsilon_p^{-1} \phi - p)_{ext}.$
Next, we assume $\varepsilon_p <<1$ and expand in powers of 
$\varepsilon_p$:
$a = a^{(0)} + \varepsilon_p a^{(1)} + O(\varepsilon_p^2)$
where $a=u_1,u_3,v_1,v_3,\phi,p$.
The equations for the displacements (\ref{adim34bis})
impose $\phi_{x_1}^0=\phi_{x_3}^0=0$.
Thus, to leading order $\phi^{0}= \phi_{\infty}$ and
$\phi = \phi_{\infty} + \varepsilon_p \phi^{(1)} + O(\varepsilon_p ).$
Inserting the expansions in Equations (\ref{adim12bis}) and (\ref{adim34bis})
and equating coefficients to 
zeroth order ($\varepsilon_p^0$)  we find
\begin{eqnarray}
v_{3,x_3}^0 =1, \quad
v_{3,x_3}^0 = (1- \phi_{\infty})^2 p_{x_3,x_3}^{(0)}, \label{eqzero12} \\
K_3 u_{1,x_3x_3}^{(0)}= p_{x_1}^{(0)} + \phi_{x_1}^{(1)}, \quad
K_2 u_{3,x_3x_3}^{(0)}= p_{x_3}^{(0)} + \phi_{x_3}^{(1)},
\label{eqzero34}
\end{eqnarray}
with boundary conditions
$(u_1^{(0)},u_3^{(0)}) \big|_{x_3=0} =(0,0),$
$u_{1,x_3}^{(0)} \big|_{x_3=1} = 0,$
$(- p^{(0)}  + 2 K_2  u_{3,x_3}^{(0)})  \big|_{x_3=1} = -p_{ext}.$
We have used $\phi_{ext}=0$ and included 
$\varepsilon_p^{-1} \phi_{\infty}$ in a term $ \varepsilon_p^{-1} p^{(-1)}$,
$\phi_\infty$ being constant the derivatives $p_{x_1},p_{x_3},p_t$ do not 
contain this term.
Equation (\ref{eqzero12}) gives $v_3^{(0)}=x_3$. Integrating in time, we get 
$u_3^{(0)}=x_3t.$ Introducing this in Equation  (\ref{eqzero34}) we obtain
\begin{eqnarray} 
p_{x_3}^{(0)}= - \phi^{(1)}_{x_3}. \label{eqzero4bis}
\end{eqnarray}

Introducing (\ref{eqzero4bis}) in (\ref{eqzero12}) we find
$1 =-(1-\phi_{\infty})^2  \phi_{x_3x_3}^{(1)},$
with boundary conditions
$\phi_{x_3}^{(1)}(1)=0$ and $\phi_{x_3}^{(1)}(0)=\phi^{(1)}(0){h \over \ell}$,
where $\ell$ is the lengthscale over which gradients develop in the substrate
due to water flowing to the biofilm, approximated by $R$. 
The latter boundary condition is established in~\cite{seminara} by matching a 
two phase flow model for the biofilm with another one for agar, we keep it
here. The solution consistent with these boundary conditions is
$ \phi^{(1)}= {1\over (1-\phi_{\infty})^2} 
\big( x_3 -{x_3^2 \over 2} + {R \over h} \big), \label{phi1}$
which gives 
$
p^{(0)}=p^{(0)}\big|_{x_3=1} +  (\phi^{(1)}\big|_{x_3=1} - \phi^{(1)} ) =  
p_{ext} + 2K_2  t  + (1-\phi_{\infty})^{-2}  \Big( {x_3^2 \over 2} -x_3  
 +{1\over 2} \Big),  
$
where the contribution $\varepsilon^{-1} = {R \over h}$ disappears thanks
to the boundary condition.  For the expansion of $\phi$ in powers of 
$\varepsilon_p$ to be consistent we need $\varepsilon_p/\varepsilon <<1$.
Notice that $u_3^{(0)}=x_3 t$ implies $u_{3,x_3}^{(0)} \big|_{x_3=1}=t$. 
To compute $u_1$ and $v_1$ we use Equation (\ref{eqzero34}), and the derivatives with respect to $x_1$ of $p$ and $\phi$, which enter this equation through the dependence on $h(x_1,t).$ To address this issue properly, we switch back to dimensional variables using $P=g h^2 \zeta$ and $x_3'= x_3/h$ to get
\begin{eqnarray}
p = {g \mu_f  \over \xi_\infty^2(1-\phi_{\infty})^2}
\Big( {x_3^2 + h^2 \over 2}- x_3 h \Big), \label{p} \\
\phi = \phi_\infty + \varepsilon_p \phi^{(1)} =
\phi_\infty + {g \mu_f  \over \Pi \xi_\infty^2 (1-\phi_{\infty})^2}
\Big( x_3 h - {x_3^2 \over 2} + R h \Big), \label{phi}
\end{eqnarray}
discarding in $p$ the lower order terms which do not contribute to the  
derivatives. 
Then, Equation (\ref{eqzero34}) yields
$\mu_s  u_{1,x_3x_3}  =  
{g \mu_f  \over \xi_\infty^2 (1-\phi_{\infty})^2} hh_{x_1}
\Big( 1 + {R\over h} \Big). $
Integrating twice and applying the boundary conditions
(\ref{bcseminara}) at $0$ and $h$ we find the displacement
\begin{eqnarray}
u_1 = {g \mu_f  \over \xi_\infty^2\mu_s
(1-\phi_{\infty})^2} R h_{x_1} \Big({x_3^2 \over 2}
- x_3 h \Big) \label{u}.
\end{eqnarray}

Differentiating the displacements $u_1$ and $u_3$ with respect to 
time we get the velocities
\begin{eqnarray}
 v_{1} =  
{g \mu_f  \over \xi_\infty^2 \mu_s
(1-\phi_\infty)^2}
\Big(  [R h_{x_1}]_t [{x_3^2 \over 2}
- x_3 h] - R h_{x_1} x_3 h_t
\Big), \quad v_{3} =x_3. \label{v}
\end{eqnarray}

Once we approximate expressions for velocities, pressures, and volume fractions are available in the $x_1x_3$ plane, the equation for the free boundary (\ref{height2}) becomes 
$
h_t + {\partial \over \partial x_1} \int_0^h  \mathbf v \cdot \hat x_1 \, dx_3 
= \mathbf v \cdot \hat x_3\big|_0, $ $
\mathbf v = \mathbf v_s - {(1-\phi)^2 \over \zeta} \nabla p,
$
where $\mathbf v$ is the volume averaged velocity 
and $\mathbf v_s$ the velocity of the solid, given by
(\ref{v}). Setting $\phi \sim \phi_\infty$, we have
$ \mathbf v \cdot \hat {\mathbf x}_3\big|_0 =  g h {(1-\phi)^2 \over 
(1-\phi_\infty)^2}\sim  g h,$
and the flux becomes
$
\int_0^h  \mathbf v \cdot \hat {\mathbf x}_1  \, dx_3 =
{ g \mu_f  \over \xi_\infty^2 \mu_s (1-\phi_\infty)^2 }
\left( - [ R h_{x_1}]_t {h^3 \over 3}  - R h_{x_1} {h^2 \over 2} h_t \right) 
- g { h^2 h_{x_1} \over 2}.
$
Performing the change of variables $h= e^{t} \tilde h$,
$h_t = e^{t} \tilde h_t +  e^{t} \tilde h$, and dropping the
symbol $\tilde{}\,$ for simplicity, we find the desired equation.

In radial coordinates the same arguments work. The equation for 
the height is then:
\begin{eqnarray*}
h_t - { g \mu_f  \over \xi_\infty^2 \mu_s
(1-\phi_\infty)^2} {1\over r}
\left( r( [ R h_{r}]_t {h^3 \over 3} +  R h_{r} {h^2 \over 2} h_t )\right)_r  
-  g {1\over r} \left( r { h^2 h_{r} \over 2} \right)_r = g h.
\end{eqnarray*}

In dimensionless variables $r= R_0 \tilde r$, $t = g^{-1} \tilde t$,
$h= h_0 \tilde h$. Dropping the symbol $\tilde{}$ for simplicity the
equation becomes
$
h_t - K {1\over r}
\left( r( [ {R\over R_0} h_{r}]_{t} h^3 +  
{R\over R_0} h_{r} {3 h^2 \over 2} h_{t} )\right)_r  
- {h_0^2 \over 2 R_0^2} {1 \over   r} \left( r h^2 h_{r}  \right)_r   =  h,
$
with
$
K={ {g \mu_f}  \over 3 \xi_\infty^2 \mu_s (1-\phi_\infty)^2 R_0}  {h_0^3}.
$
Assuming  $\epsilon={h_0/R_0} << K$ we get
\begin{eqnarray}
h_t - K {R_t\over R_0} {1\over r} \left( r h_{r} h^3  \right)_r
- K {R\over R_0} {1\over r} \left( r  h_{rt}  h^3 \right)_r
- K {R\over R_0} {1\over r} \left(r h_{r} {3 h^2 \over 2} h_{t} \right)_r   =  h.
\label{h}
\end{eqnarray}

Performing the change of variables $h= e^{t} \tilde h$,
$h_t = e^{t} \tilde h_t +  e^{t} \tilde h$, and dropping again the
symbol $\tilde{}\,$ for simplicity, we find (\ref{echeight}).

\end{document}